\documentclass[runningheads]{llncs}
\usepackage[T1]{fontenc}
\usepackage{graphicx}

\usepackage{tikz}
\usepackage{amsmath}
\usepackage{hyperref}
\usepackage{etoolbox}
\usepackage{array}
\usepackage{algorithm}
\usepackage{algorithmic}

\usepackage{listings}
\usepackage{tcolorbox}
\tcbuselibrary{listings,skins}

\usepackage{listings}
\usepackage{xcolor}
\begin{document}
\title{Firewall3D: A Hardware Firewall for Defending 3D Printers Against Firmware Attacks}
\titlerunning{Firewall3D}
%


\author{Seyed Ali Ghazi Asgar
\and
Narasimha Reddy}

\authorrunning{SA. Ghazi Asgar and N.Reddy}
%
\institute{Texas A\&M University, Department of Electrical and Computer Engineering, College Station,TX,USA\\
\email{\{alighazi,reddy\}@tamu.edu}}

\maketitle              
\begin{abstract}
As the 3D printing market continues to grow rapidly, with an estimated value exceeding \$30~billion, cybersecurity risks and attacks targeting additive manufacturing systems are also increasing. These attacks aim to sabotage printed components, steal intellectual property, or even physically damage the 3D printer itself. One major cybersecurity threat in this domain is firmware level attacks, which can be introduced through supply chain compromises, malicious firmware updates, or insider threats that deploy modified firmware to manipulate printer behavior. To defend against such threats, we propose a dedicated hardware based security solution, \textit{Firewall3D}, that acts as a hardware firewall for 3D printers. \textit{Firewall3D} continuously monitors physical layer signals, including stepper motor currents, end stop switches, nozzle and bed temperatures and cooling fans, to verify that the printer’s physical behavior matches the intended G-code execution. Our experimental results demonstrate that \textit{Firewall3D} can effectively detect a wide range of firmware attacks that could compromise print integrity, damage printer components, or leak intellectual property. Upon detecting abnormal behavior, the system can immediately trigger an alarm and halt the printing process, thereby preventing further damage and risks.

\end{abstract}
\section{Introduction}
3D printing is an additive manufacturing (AM) technique that creates objects by stacking successive layers of material on top of one another. Although AM was initially used primarily for prototyping, it is now widely employed to produce parts with complex geometries \cite{gibson2021additive,tofail2018additive,debroy2018additive,moneysonigara2025point,marketsandmarketsGlobalPrinting}. For example, recently, the Apple Watch Ultra 3 was manufactured using 3D printing technology using recycled aerospace grade titanium powder \cite{apple}. 

Additive manufacturing is  subject to various threats. In particular, it faces numerous cybersecurity challenges. One major concern in this area is the theft of intellectual property (IP)\cite{thakare2025secure,ahsan20253d,shomenov2025cost,moulika2024optimizing,yuan2024materials,puch2025securing,asgar2025never,kumar2025securing,elgez2025intellectual,mishra2025real,asgar2026quietprint,jamaranil2025practitioner}. In such scenarios, attackers attempt to steal design files or G-code files in order to manufacture counterfeit components, which often lack the strength, quality, and reliability of the original parts 

To keep design files secure, researchers in \cite{baumann2017modelstream,tiwari2020cybersecuritystream} proposed a streaming based solution. In this approach, the client streams the G-code files line by line to the manufacturing site, and each command is deleted from memory immediately after execution, keeping the G-code file safe. However, this approach raises security concerns for manufacturers. G-code files can be dangerous or compromised \cite{beckwith2021needle,ali2025machine,rossel2025security}, and executing them on the manufacturing site can potentially damage expensive machinery. To address this issue, the authors in \cite{asgar2025never} proposed a novel design file streaming method. In their approach, the design file is first sliced into horizontal layers and then sent to the manufacturer. A real-time design-file-to G-code translator on the manufacturing side converts these layers into G-code commands. Since the G-codes are generated locally at the manufacturing site, they are safe and do not pose a risk of damaging the machine. Therefore,  the manufacturer and the client are both protected.  

In addition to targeting design files, attackers may also exploit 3D printer hardware. In side-channel attacks, adversaries use indirect signals emitted by the machine to infer or reconstruct the printed objects.  Various side channels have been exploited to infer sensitive information from 3D printers. In the optical domain, cameras used for monitoring or anomaly detection can be compromised which allows attackers to reconstruct G-code from visual data, as demonstrated in \cite{chattopadhyay2025oneoptical}. Power based side-channel attacks have also been explored, where current measurements from stepper motors were analyzed to detect motion states and reconstruct printing instructions \cite{power_attack}. Vibration based approaches utilize accelerometer data to capture motor vibrations, from which spectrogram analysis can reveal nozzle movements and enable reconstruction of printed objects \cite{vibration}. Furthermore, magnetic side-channel attacks exploit electromagnetic emissions from stepper motors; studies such as \cite{magneticjamarani2025practitioner,magneticsong2016my} showed that even a nearby mobile phone can capture magnetic signals sufficient to reconstruct manufactured objects. QuietPrint \cite{asgar2026quietprint} investigates acoustic side-channel leakage in 3D printers and demonstrates how signals from cooling fans and stepper motors can be exploited to reconstruct printed objects. To mitigate this risk, it proposes the Stealth Head Movement (SHM) algorithm that obfuscates motion patterns.

Besides side-channel attacks, firmware attacks represent another significant threat vector. Marlin firmware is one of the most widely used open-source firmware platforms for 3D printers. Moore et al.~\cite{moore2016vulnerability,moore2017implications} identified a buffer overflow vulnerability as well as exploitable communication channels between the PC and the 3D printer. By altering the G-code sent from the PC to the 3D printer, they were able to compromise the structural integrity of printed objects.  In another study, the researchers proposed ten different firmware attacks to compromise a bio 3D printer. They successfully modified the firmware to manipulate print metadata, overheat stepper motors, damage the nozzle, degrade air quality, launch thermal attacks, manipulate printer speed, and cause under extrusion and over-extrusion~\cite{bioprint}. \textit{FLAW3D} is another firmware based attack proposed by Pearce et al.~\cite{flaw3d}. In this work, attackers manipulated the UART interrupt service routine of the 3D printer microcontroller to change benign G-code commands into malicious ones, reducing the strength of the printed object by up to 50\%. Another work proposed by \cite{basu2026scream} used an invasive approach to monitor the signal between the microcontroller and the stepper motor driver. Although their approach achieves performance similar to our system, it requires soldering and is therefore not generalizable across different 3D printers. In contrast, our proposed approach captures the signal after the motor driver, making it more generalizable across different motherboards and different 3D printers.

The main focus of our work is to propose a comprehensive defense mechanism to safeguard 3D printers against firmware attacks. To this end, we design the first dedicated hardware firewall for 3D printers. Our custom designed hardware, named \textit{Firewall3D}, monitors physical-layer signals to distinguish between benign and malicious behavior. \textit{Firewall3D} captures stepper motor signals, temperature readings ,and end stop switch signals in real time. When any mismatch is detected between the expected values and the observed signals, an alert is triggered, and appropriate mitigation strategies are executed. The main contributions of this paper are summarized as follows:
\vspace{-4pt}
\begin{itemize}
    \item Propose, design and validate an approach to detect and defend against firmware attacks
    \item Built a hardware firewall for 3D printers to realize the design in practice 
    \item Real time physical layer signal monitoring and anomaly detection
    \item Experimental validation against attack scenarios
\end{itemize}

The remainder of this paper is organized as follows. Sections~\ref{sec:terminology} and \ref{sec:ThreatModel} introduce general terminology used in the 3D printing domain and describe the threat model respectively. Sections~\ref{sec:hardware} and  \ref{sec:hardwareEval} presents the design and implementation of the proposed Firewall3D. Section~\ref{sec:attackAndDefense} showcases attack scenarios and corresponding defense mechanisms. Finally, Section~\ref{sec:concl} concludes the paper.

\section{3D printing terminology}
\label{sec:terminology}
\subsubsection{Computer Aided Design (CAD):}
The first stage of manufacturing a product begins with designing the component. Engineers use computer aided design (CAD) applications to model three dimensional objects. Each CAD software stores design files in its own proprietary format. Therefore, it is necessary to convert these files into a universal format that can be widely used across different platforms and tools.
\vspace{-15pt}
\subsubsection{Standard Triangle Language (STL):}
The STL format is one of the most widely used file formats in the additive manufacturing domain. An STL design file represents an object as a collection of triangular facets, each defined by the coordinates of its vertices. These triangles collectively form a mesh that creates the geometry of the original design.
\vspace{-15pt}
\subsubsection{G-code format:}
STL files cannot be executed directly by a 3D printer. Therefore, an STL file is first converted into machine executable instructions, known as G-code, by the slicer application. The slicer uses various parameters such as print temperature, print location, number of objects, printing speed, material type, and layer thickness to generate the G-code commands. Once generated, these commands are executed line by line by the printer. For instance, the \texttt{G0} and \texttt{G1} commands are used to move the printer’s nozzle to a specified location, depending on whether material extrusion is required or the nozzle temperature can be set to 200~$^\circ$C using the \texttt{M104 S200} command.


\subsection{3D printer hardware}
\begin{figure}[!h]
    \centering
    \includegraphics[width=0.65\linewidth]{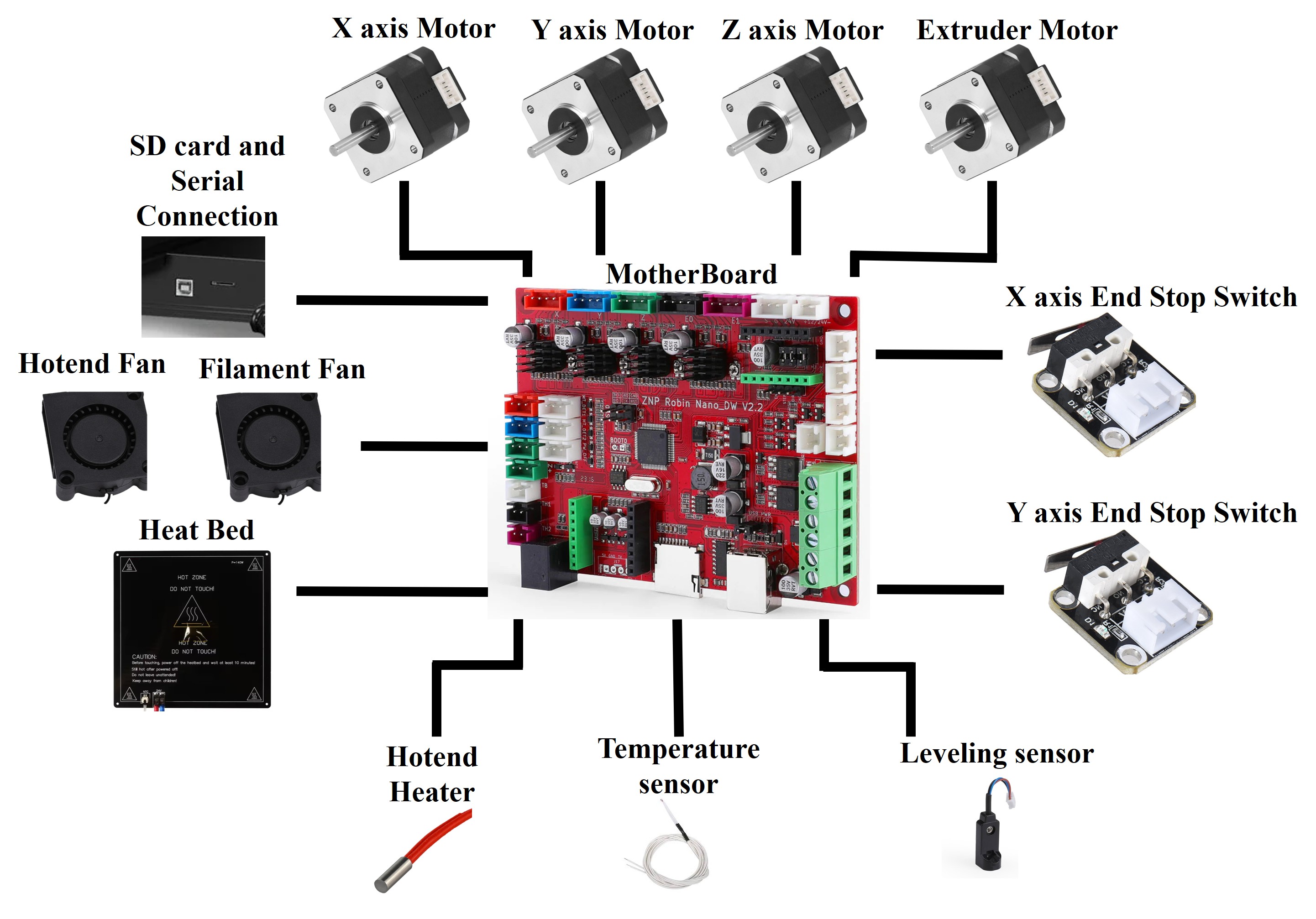}
    \caption{Overview of sensors and actuators in a commercial 3D printer connected to the main motherboard. }
    \label{fig:printerHardware}
\end{figure}

In this subsection, we describe the sensors and actuators commonly found in a commercial 3D printer setup. A brief overview of these components is shown in Figure~\ref{fig:printerHardware}. All components are connected to a main motherboard, which controls the actuators and sensors according to the 3D printer’s firmware. In this work, our 3D printer uses Marlin firmware, one of the most widely adopted open-source firmware platforms in the 3D printing community.

In total, the system consists of nine types of sensors and actuators: stepper motors, cooling fans, an SD card reader, a serial communication interface, a heated bed, a hotend heater, temperature sensors, bed-leveling sensors, and endstop switches. We discuss each of these components in detail in the following sections.

\vspace{-15pt}

\subsubsection{Stepper Motors:}
The movement of the 3D printer nozzle and the extrusion of filament are controlled by stepper motors. In Cartesian 3D printers, a dedicated stepper motor is used for each axis. In total, four stepper motors are typically employed: three for motion along the X, Y, and Z axes, and one for driving the filament extrusion mechanism.
\vspace{-10pt}

\subsubsection{End Stop Switches and Leveling Sensor:}
Stepper motors can be used to determine relative changes in movement; however, they do not provide absolute position information. Therefore, it is necessary to define a reference zero point and calibrate motion relative to that coordinate. To achieve this, each axis of a 3D printer is equipped with an endstop switch. During the homing process, the stepper motor moves in an open-loop manner until the corresponding endstop switch is triggered. Once activated, the printer sets the current nozzle location as the home position. Because the Z-axis is more delicate and precise contact with the build plate is critical, more accurate sensors such as force sensors or proximity sensors are often used to detect the point at which the nozzle tip touches the build plate. If the Z-axis offset is not properly calibrated, the initial layers may not adhere well to the build plate, leading to print failure.

\subsubsection{Fans:}
Two types of fans are commonly used in 3D printers. The first fan is used for cooling the hotend and maintaining its temperature within the desired operating range. The second fan is directed toward the build plate and is used to cool and solidify the molten filament immediately after it exits the hotend.

\subsubsection{Heat Bed:}
The build plate of a 3D printer is equipped with heating elements. During the printing process, it is important to maintain the build plate at a  high temperature to ensure proper adhesion of the initial filament layers. A temperature sensor is also embedded in the build plate to continuously monitor its temperature.
\vspace{-10pt}

\subsubsection{Hotend Heater and Temperature Sensor:}
As shown in Figure~\ref{fig:printerHardware}, the hotend of a 3D printer includes two primary active components. The first is the hotend heater, also known as the hotend cartridge, which is responsible for raising the temperature and melting the filament. A temperature sensor is placed adjacent to the hotend cartridge to provide temperature readings to the firmware, which is used for a closed loop temperature control through a PID controller. 
\vspace{-10pt}

\subsubsection{Serial Connection and SD card interface}
Most 3D printers are equipped with two primary communication interfaces. One common interface is an SD card reader, which allows the printer to read G-code files directly from removable storage. The second interface is a USB Type-B port connected to a USB-to-serial converter interface. This port allows a host computer to send serial commands and G-code instructions directly to the printer.


\begin{figure*}[!t]
    \centering
    \includegraphics[width=0.8\linewidth]{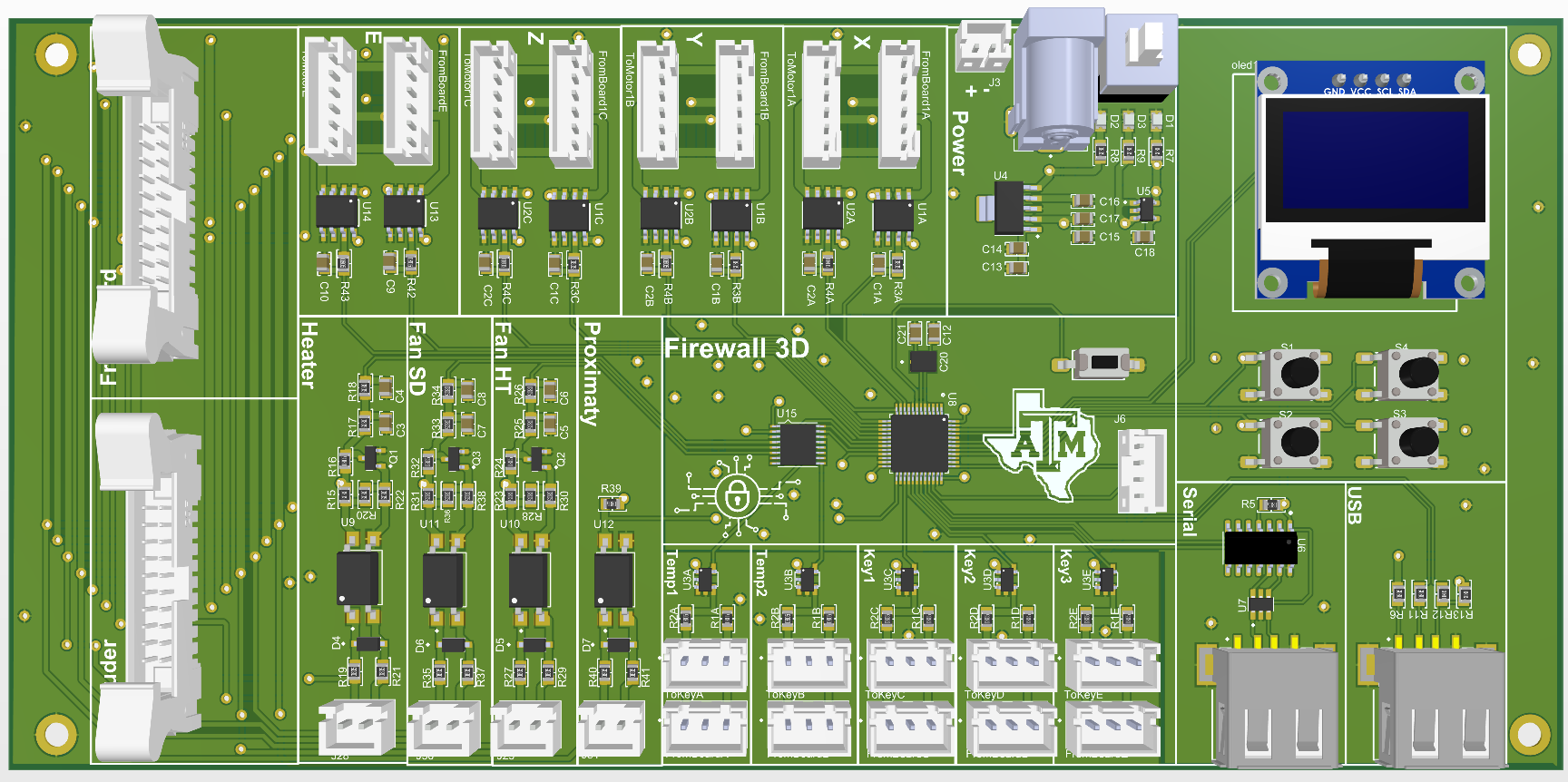}
    \caption{Firewall3D PCB overview.  }
    \label{fig:mainBoard}
\end{figure*}

\section{Threat Model}
\label{sec:ThreatModel}
In this work, our primary focus is on detecting hardware level attacks. We assume that the main motherboard may be compromised through supply chain attacks, hardware Trojan insertion, or the deployment of maliciously modified firmware. The attacker’s objectives include compromising the mechanical strength of the printed object, leaking the design’s intellectual property through side-channel mechanisms or even damaging the 3D printer machine. 

To defend against these threats, we propose a dedicated hardware-based solution that monitors physical layer signals including motion commands, temperature sensor readings, fan speed values, and other critical signals in real time. By independently observing these signals, the proposed system mitigates the risks of intellectual property theft, component sabotage, and machine damage. 

We assume physical access to the 3D printer hardware during defense method. Notably, our approach does not require any modifications to the printer’s motherboard or firmware, as it operates using only minimal additional wiring.

\section{Firewall3D hardware}
\label{sec:hardware}

\subsection{Our Approach}
\begin{figure}[!h]
    \centering
    \includegraphics[width=1\linewidth]{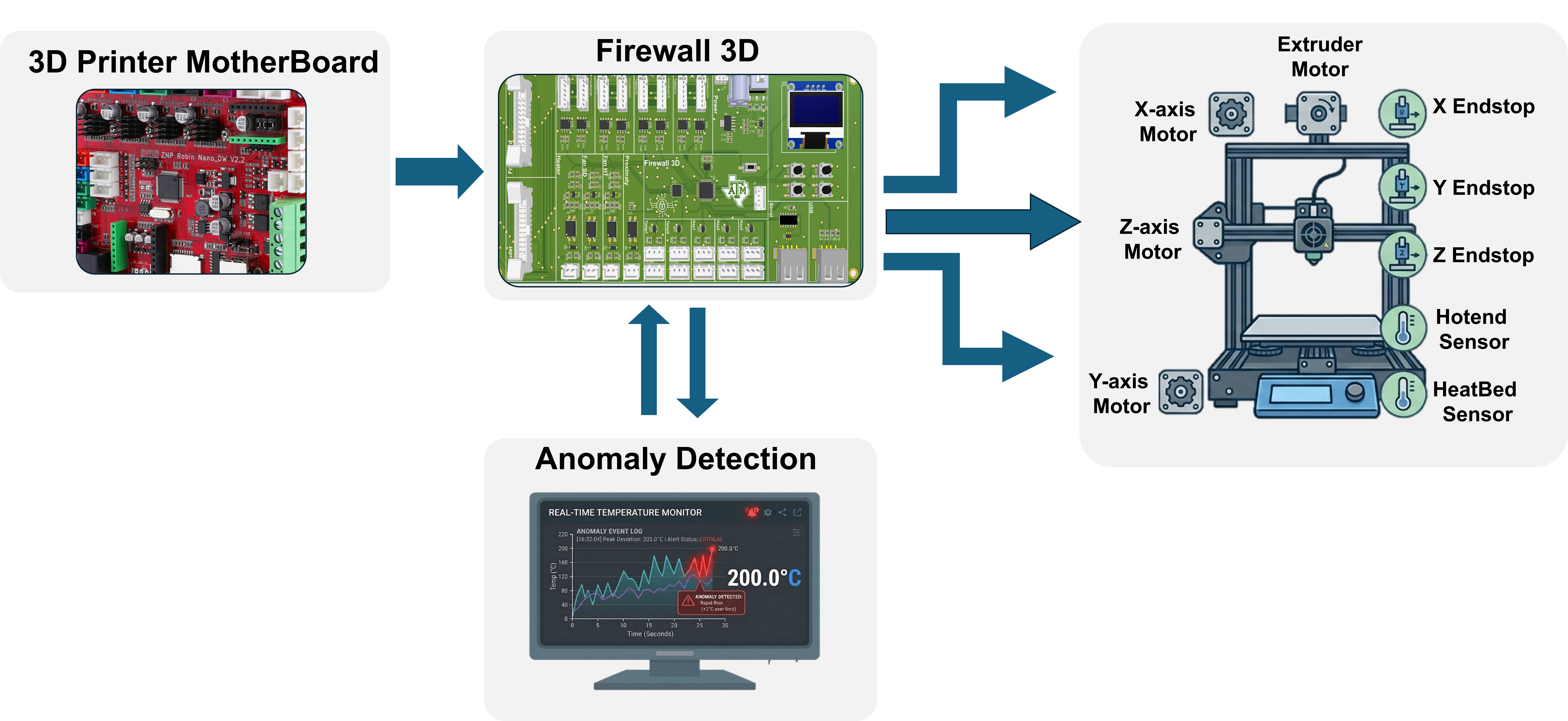}
    \caption{Overview of the Proposed Approach to detect firmware attacks in real time. }
    \label{fig:ourapproach}
\end{figure}

The primary concept of our work can be understood through an example: our hardware functions as a monitoring checkpoint, similar to a security system that inspects every communication entering a network. We have developed a device called Firewall3D that operates between the printer's motherboard and its sensors/actuators which enables us to  monitor  all signals that pass through the system (See Figure~\ref{fig:ourapproach}). The fundamental principle is a "bump in the wire" approach: rather than the motherboard sending commands directly to components such as stepper motors, all communications are routed through our Firewall3D first. This arrangement allows us to examine each command and identify any anomalous behavior. For instance, when the motherboard issues a command to the stepper motor, the signal passes through our device before reaching the motor which provides an opportunity to validate the signal. Should any irregularity be detected, our system can immediately identify and flag it. 
Our approach requires only additional wiring to connect the motherboard to Firewall3D and then to the target actuators and sensors.

In this section, we provide an overview of the components of our hardware designed to secure the 3D printer. Figure~\ref{fig:mainBoard} shows the \textit{Firewall3D} PCB. The system is centered around an STM32F103C8K6 microcontroller, which manages signal acquisition via the ADC, communication over USB and serial interfaces, I\textsuperscript{2}C communication for the OLED display, and user interactions through keys.
\vspace{-10pt}
\subsection{Fan Speed Measurement}
Fans are vital components in a 3D printer, as they regulate temperature and cool the filament. Therefore, it is important to monitor their signals to ensure that fan speed is not tampered with. Fan speed is  controlled using Pulse Width Modulation (PWM). To estimate the actual speed, it is necessary to measure the PWM signal. 

\begin{figure}[!h]
    \centering
    \includegraphics[width=0.7\linewidth]{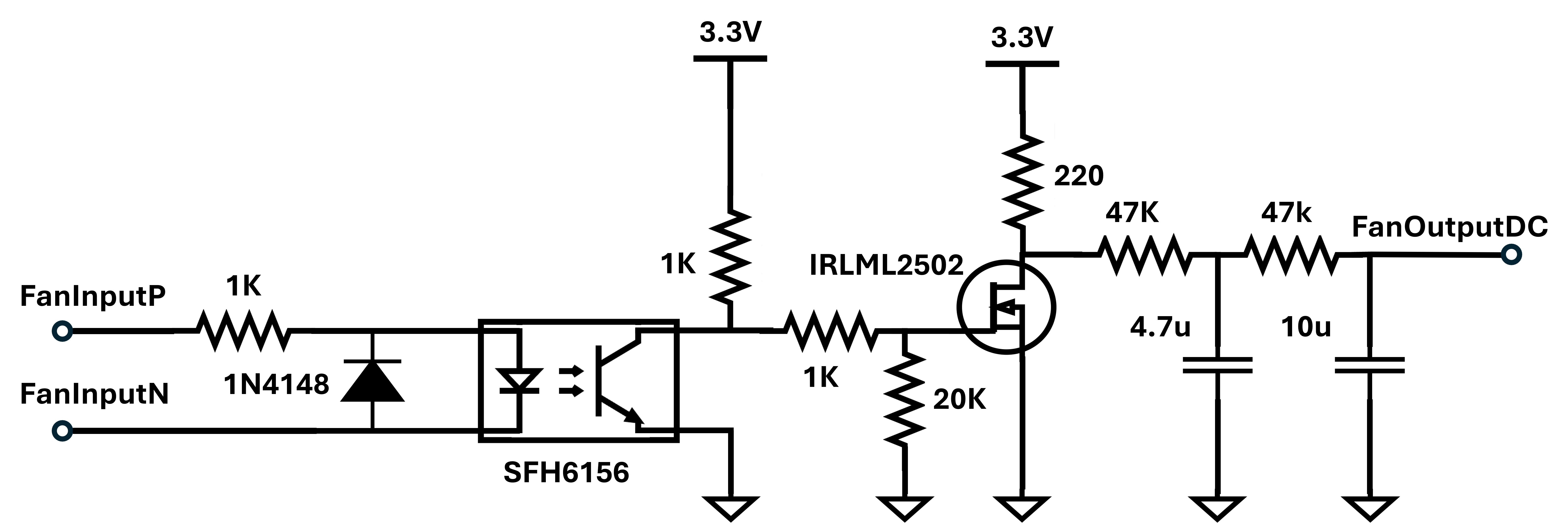}
    \caption{Proposed circuit for converting the fan PWM signal into a DC voltage. The resulting DC signal is then read by the microcontroller's ADC.}
    \label{fig:fanCircuit}
\end{figure}

To achieve this, we propose the circuit shown in Figure~\ref{fig:fanCircuit} to convert PWM signals into DC voltages that can be directly read by the microcontroller's analog-to-digital converter (ADC). As illustrated, we first use an SFH6156 optocoupler to electrically isolate the circuit, ensuring that the fan circuit and 3D printer motherboard remain unaffected. A second order low-pass cascade filter is then used to extract the DC voltage corresponding to the PWM duty cycle. 
Since the optocoupler cannot drive the low-pass filter directly, we employ an IRLM2502 NMOS transistor as an inverting buffer to drive the filter. We simulated this circuit using different PWM duty cycle values at a frequency of 7.8~Hz, which corresponds to the frequency used by our 3D printer. The simulation result is also provided in the Appendix section \ref{sec:coolfansimu}.

\vspace{-15pt}

\subsection{Heat Bed Temperature Measurement}
The heated bed is responsible for raising and maintaining the build plate temperature. Since an attacker may manipulate the firmware to alter this temperature, it is necessary to monitor the PWM signal driving the heated bed. To achieve this, we reuse the same PWM-to-DC conversion circuit shown in Figure~\ref{fig:fanCircuit} to measure the heated bed’s PWM duty cycle. Because the PWM frequency of the heated bed is lower than that of the cooling fans, we adjust the RC filter values accordingly to accommodate this difference.
\vspace{-10pt}

\subsection{HotEnd Temperature Measurement}
\begin{figure}[!h]
    \centering
    \includegraphics[width=0.45\linewidth]{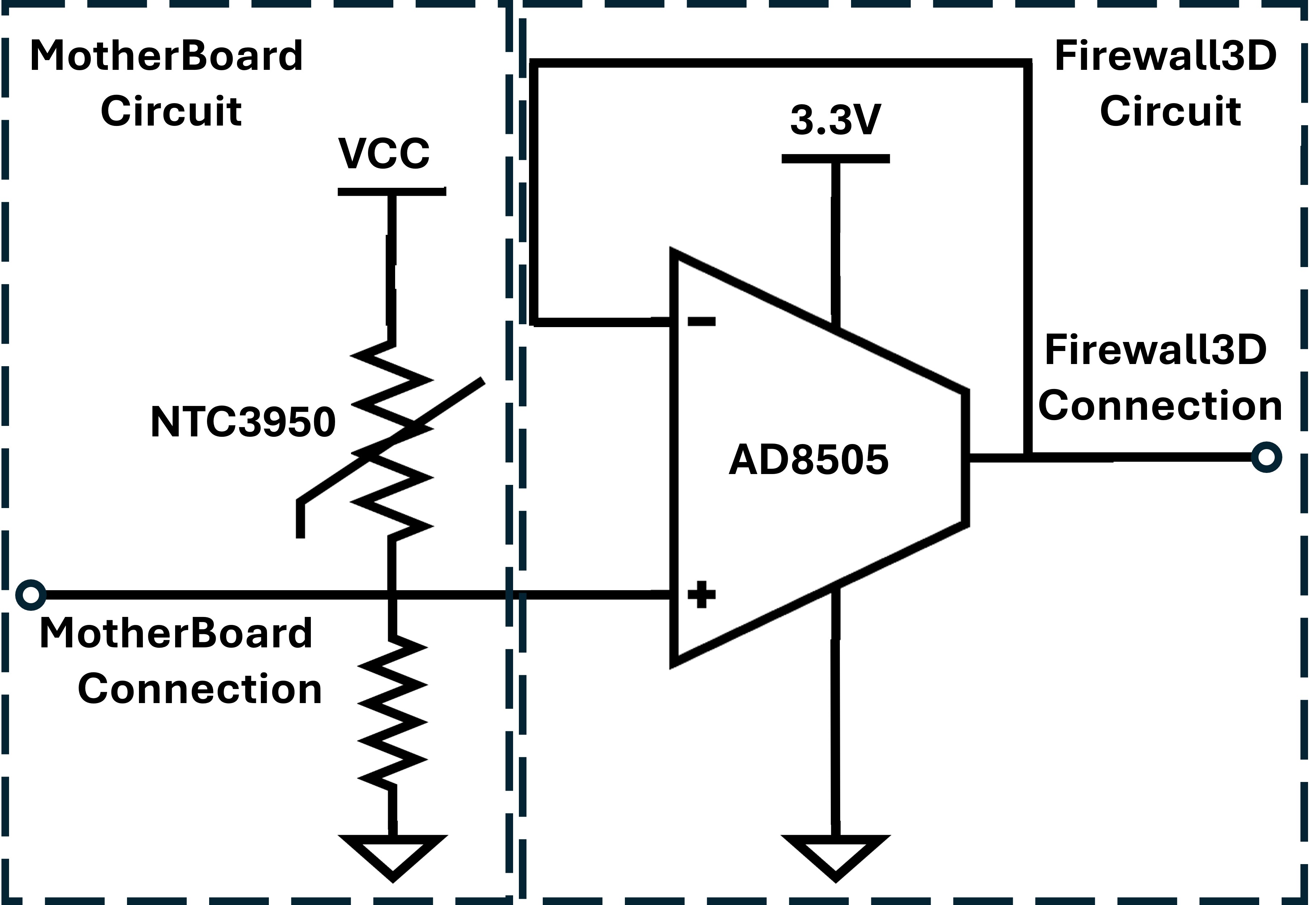}
    \caption{Circuit for acquiring the hotend temperature. The pull-down resistor configuration is already present on the mainboard; therefore, only a buffer circuit is added to enable non-intrusive temperature measurement.}
    \label{fig:tempCircuit}
\end{figure}

The 3D printer used in this work uses an NTC3950 temperature sensor. This sensor resistance varies with temperature. Therefore, by using a pull-down resistor configuration, the  voltage divider output can be measured and converted back into temperature readings. To acquire data from this sensor, we use the circuit shown in Figure~\ref{fig:tempCircuit}. Since we do not want to interfere with the main board’s reading of this sensor, we place a buffer after the sensor to measure the voltage from the circuit and forward it to the ADC pin of our microcontroller.

\vspace{-30pt}

\subsection{End Stop Keys Monitoring}
The same buffer configuration used for the temperature measurement (Figure~\ref{fig:tempCircuit}) is also applied to the end stop switches. When a switch is pressed, the microcontroller reads a 3.3~V signal and when the switch is released, it reads a logical high signal of 0~V.


\subsection{Computer Integration}
Since \textit{Firewall3D} is designed as a dedicated protection platform, it is necessary to incorporate additional features that support future development and system flexibility. For example, instead of interfacing an SD card directly with the microcontroller, we integrate a separate computer  to handle SD card access. This design choice enables the execution of custom scripts to analyze and verify the sanity of G-code commands before they are forwarded to the printer.

Although such checks could be implemented on the microcontroller, offloading high level analysis to a computer provides greater scalability, simplifies software updates, and facilitates the deployment of advanced safety and security policies. This hybrid architecture allows the microcontroller to focus on real-time signal monitoring, while the computer executes computationally intensive and high level verification. Therefore, we added a dedicated USB to serial converter (MCP2221A) to enable serial communication with the PC.
For instance, motor current decoding into motion length and speed is handled on the microcontroller side, as stepper motor signals have a high sampling rate and transmitting raw data to the PC could introduce additional delays. In contrast, temperature and fan measurements are sent to the PC, where ADC values are converted into corresponding physical quantities, since sensor specific parameters from different sensors are more conveniently configured on PC.

Moreover, in this work, we used a laptop to simulate firmware-level attacks. As an example, consider a scenario in which the original G-code command intended for execution is \texttt{G0 X10}. If the printer firmware is compromised, this command may be altered to \texttt{G0 X9.9}, resulting in subtle positional deviations that can accumulate and degrade the dimensional accuracy of the printed object. To emulate this attack scenario without modifying or recompiling the 3D printer’s motherboard firmware, we transmit the modified command (\texttt{G0 X9.9}) to the printer for execution. This approach enables the controlled simulation of firmware manipulation attacks by injecting compromised commands directly from the computer to the printer while sending the original command to  the \textit{Firewall3D} PCB. By monitoring physical layer signals and comparing the observed behavior against the expected command, \textit{Firewall3D} can detect discrepancies between the intended and executed motions. Upon identifying such deviations, the system can trigger an alarm and indicate a potential firmware compromise. Firewall3d sends all the signal back to the computer and then it is possible to watch all signals in a dashboard on the PC GUI interface.

 \vspace{-10pt}

\subsection{Stepper Motor Current Measurement}

To monitor motion, it is necessary to observe the signals driving the stepper motors. Each stepper motor consists of two coils, and the current flowing through each coil must be measured separately to determine the direction of motion. To achieve this, we use two TMCS1107A3B current sensors per motor. The TMCS1107A3B is a Hall effect current sensor capable of measuring both AC and DC currents. The input current passes through a 1.8~m$\Omega$ internal conductor, generating a magnetic field that is sensed by the Hall-effect element. With a sensitivity of 200~mV/A, the sensor provides sufficient dynamic range to accurately measure stepper motor currents.
\begin{figure}[!h]
    \centering
    \includegraphics[width=0.5\linewidth]{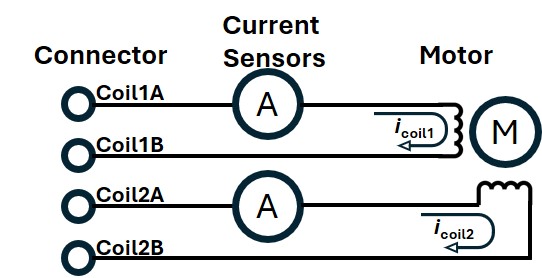}
    \caption{Stepper motor current measurement design.  }
    \label{fig:sche_stepper}
\end{figure}
As shown in Figure~\ref{fig:sche_stepper}, our setup includes two connectors. The first connector interfaces with the 3D printer’s motherboard, which provides the stepper motor drive signals. The current measurement sensors are placed inline with these signals, and the measured currents are then routed through a second connector directly connected to the stepper motor.

Since our system includes four stepper motors corresponding to the X, Y, and Z axes and the extrusion mechanism. We replicated this setup four times, as shown in Figure~\ref{fig:mainBoard}. The output of each current measurement IC is passed through a first order RC low-pass filter to attenuate high-frequency noise and further smooth the signal for better ADC measurements.

\subsection{Sampling Analog Signals}
To sample the stepper motor current measurements, we configure a  timer on the microcontroller to generate a trigger signal at the frequency of 4~kHz. At each trigger event, an ADC conversion is initiated, and all eight ADC channels corresponding to the two coils of each of the four stepper motors are sampled sequentially. The conversion results are transferred directly to memory using the Direct Memory Access (DMA) unit. Upon completion of the sampling process, an interrupt callback is generated to signal that the data is ready.
Since the microcontroller does not provide a sufficient number of ADC input channels, we used an ADG708BRUZ analog multiplexer to acquire additional signals, including the hotend temperature, heat bed temperature, and fan monitoring signals.

\subsection{Decoding Stepper Motor Current Signals to Nozzle Position}

\begin{figure*}[!t]
    \centering
    \includegraphics[width=1\linewidth]{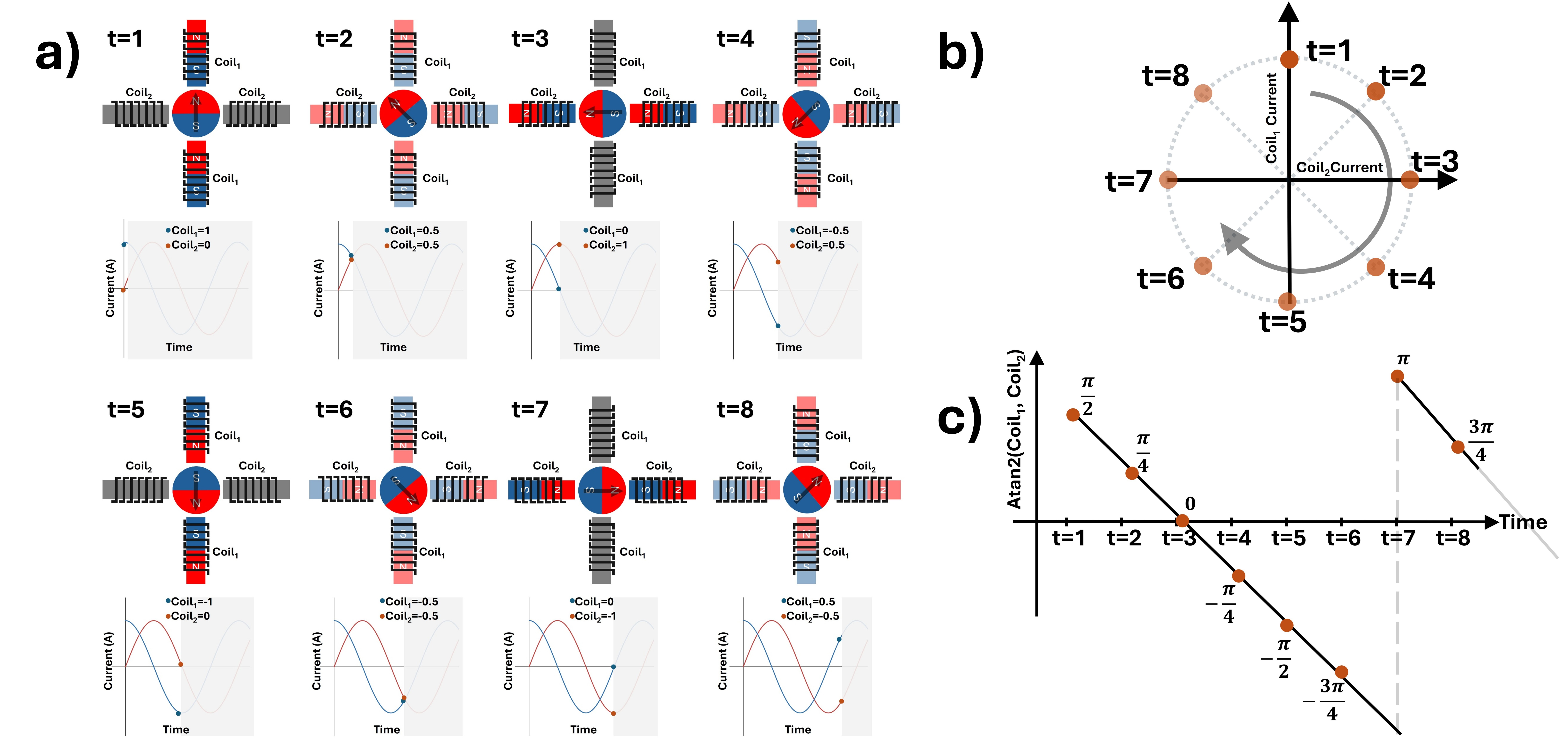}
    \caption{Decoding stepper motor current signals to infer motion. 
    (a) Simplified illustration of stepper motor operation showing the periodic excitation of two coils and the resulting rotor positions during a rotation cycle. 
    (b) Representation of the measured coil currents, where the current of the first coil is plotted against the second, revealing a circular trajectory corresponding to rotational motion. 
    (c) Angular position extracted using the $\texttt{atan2}(\textit{coil}_1,\textit{coil}_2)$ function over time, demonstrating a linear change in angle that can be converted into linear displacement.}
    \label{fig:stepperDecode}
\end{figure*}

To monitor the motion of the 3D printer, it is necessary to convert the observed stepper motor current signals into positional information and subsequently reconstruct the executed G-code commands. This enables comparison between the executed commands and the expected G-code values.

As mentioned earlier, a stepper motor consists of two coils, each of which is excited periodically to produce rotational motion. Figure~\ref{fig:stepperDecode}(a) illustrates a simplified stepper motor operation with a two-pole rotor. At time $t=1$, one coil is energized, causing the rotor to align with the corresponding magnetic field. At the next step, both coils are energized with equal current, moving the rotor to a position between the two coils. This excitation sequence continues until a full rotation is completed, after which the cycle repeats.

To convert the measured coil currents into positional information, we plot the current of the first coil versus the current of the second coil, as shown in Figure~\ref{fig:stepperDecode}(b). This representation reveals a circular (rotational) trajectory corresponding to the motor’s rotation. By computing the phase angle using the $\texttt{atan2}(\textit{coil}_1, \textit{coil}_2)$ function over time (Figure~\ref{fig:stepperDecode}(c)), we obtain a linearly changing angular signal that represents the motor’s rotational steps.

It is important to note that the range of the $\texttt{atan2}$ function is $(-\pi, \pi]$, which introduces a discontinuity at $-\pi$. To convert this angular signal into linear displacement, we count the number of angular increments of $\pi/4$ and scale the result using a printer-specific conversion factor. The \texttt{peak\_counter} variable counts the number of increments and is reset after the current motion stops and before the next motion begins.
For the 3D printer used in this work, every ten $\pi/4$ increments correspond to a linear displacement of 0.1\,mm. Therefore, the accumulated count is divided by 100 to obtain the nozzle displacement in millimeters (see Algorithm~\ref{alg:update_stepper}).

\vspace{-10pt}
\subsubsection{Stepper Motor Motion Start/Stop Detection:}
\label{sec:sterpperStopStart}
It is important to determine when to start and stop counting the \texttt{peak\_counter} variable in order to know precisely when a specific G-code motion begins and ends. To achieve this, we used a specific G-code command that toggles a dedicated pin whenever the current G-code execution is completed. In our setup, we use the pin that normally drives the 3D printer's LED. This pin is connected to our hardware through an optocoupler, which let us safely monitor the start and end of each motion without interfering with the printer’s motherboard (Please refer to Appendix Listing~\ref{lis:gcode_toggle} for more detail). 

\vspace{-1pt}

\section{Experimental Evaluation }
\label{sec:hardwareEval}
In this section, we present the real-world experimental evaluation of the hardware setup, analyzing each component individually.

\subsection{Fan Speed Measurements}
\begin{figure}[!b]
    \centering

    \includegraphics[width=0.5\linewidth]{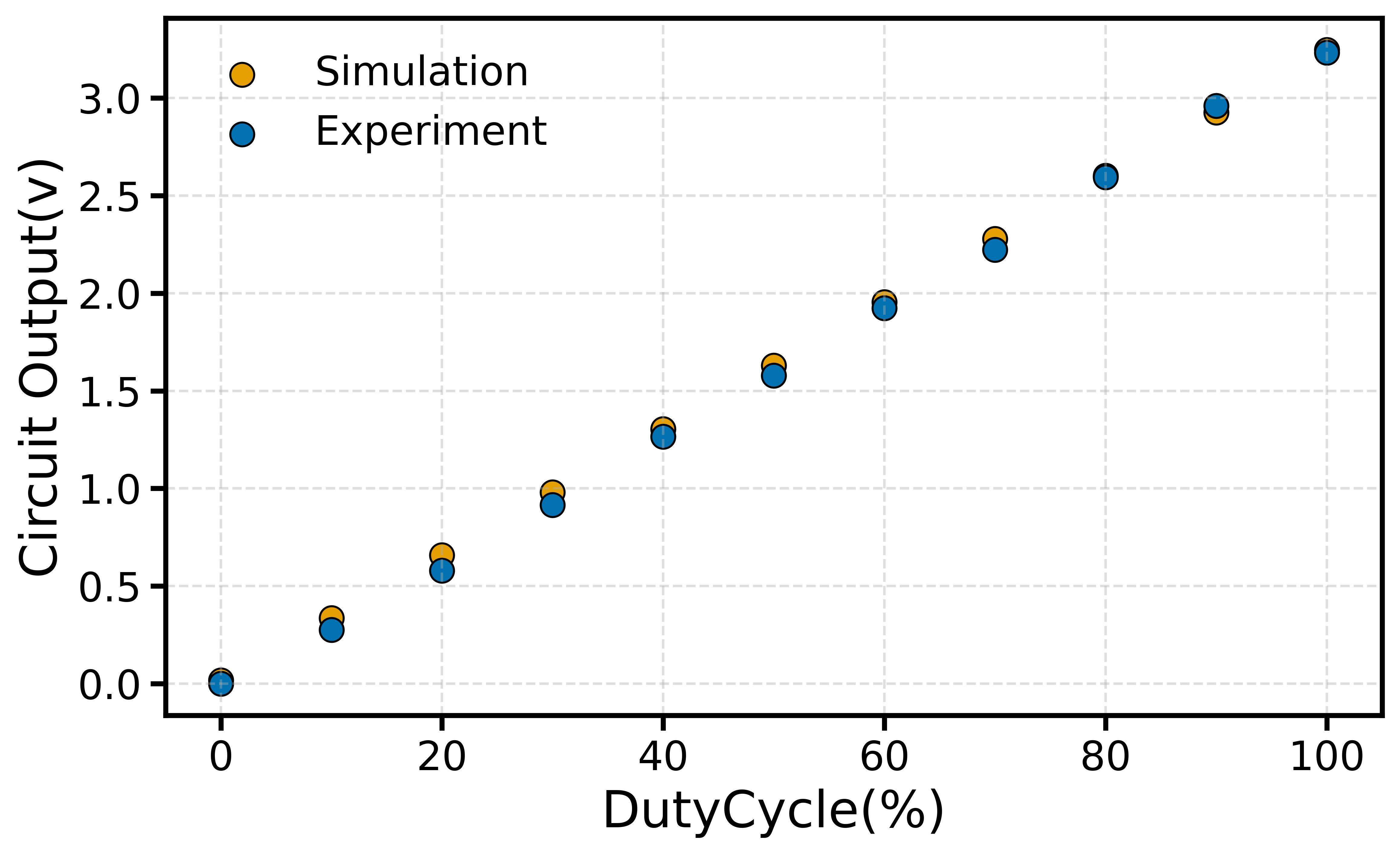}
    \caption{The output of the duty cycle to DC voltage circuit.}

    \label{fig:fanperformance}
\end{figure}

To evaluate the performance of our fan speed measurement circuit, we created a G-code program that increments the fan speed by 10\% every 10 seconds. The resulting measurements are shown in Figure~\ref{fig:fanperformance}. As illustrated in the plot, the acquired voltage can be accurately converted into the corresponding fan speed, that proves the effectiveness of our PWM-to-DC conversion circuit.

\subsection{Hotend Temperature Measurements}
\begin{figure}[!h]
    \centering

    \includegraphics[width=0.5\linewidth]{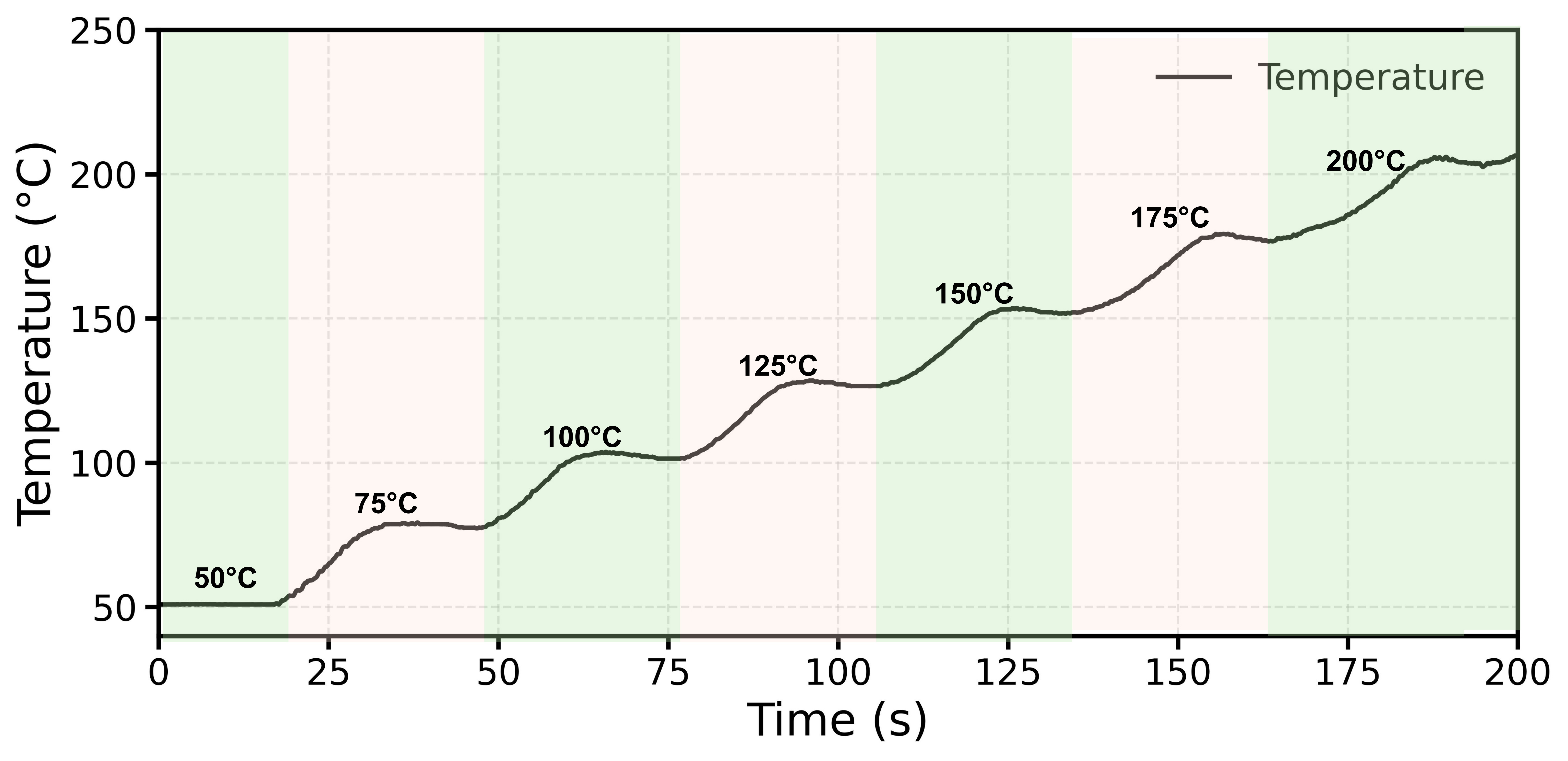}

\caption{Performance evaluation of the temperature measurement system. Color transitions denote the initiation of successive temperature setpoints. The temperature is increased in increments of 25\,°C, with each setpoint held constant for 30 seconds.}

    \label{fig:tempPerformance}
\end{figure}

To evaluate the performance of our temperature measurement circuit, we used a custom G-code program to increase the temperature by 50~\textdegree C starting from 50~\textdegree C to 200~\textdegree C. The results are plotted in Figure~\ref{fig:tempPerformance}.

\subsection{End Stop Keys Monitoring}
The output signal of our endstop switch circuit is shown in Figure~\ref{fig:stopswithcattack}. As depicted in the plot, each actuation of the endstop switch produces a corresponding change in the digital output signal. The calibration process for each axis involves two consecutive contacts with the endstop switch. Specifically, the switch transitions through the following sequence: initially OFF, then ON upon first contact (See Figure~\ref{fig:stopswithcattack}). The stepper motor then retracts slightly, causing the switch to return to OFF. Subsequently, the motor advances again to trigger the switch to the ON state. Therefore, each complete calibration cycle exhibits an ON-OFF-ON-OFF pattern, establishing a reliable homing reference for the axis.

\vspace{-10pt}
\subsection{Stepper Motor Position Estimation}

To evaluate the performance of our current position estimation hardware and algorithm, we designed a custom G-code test program. The nozzle was initially moved to the center position, then displaced to $+1~\mathrm{mm}$, $-1~\mathrm{mm}$, and returned to the center. This procedure was repeated with progressively larger increments of $10~\mathrm{mm}$, moving the nozzle to $+10~\mathrm{mm}$, then $-20~\mathrm{mm}$, and back to the center, continuing in this manner up to $100~\mathrm{mm}$. The experiment was conducted at different feed rates: F500, F1000, F1500, and F2000. As shown in Table~\ref{tab:feedrate_error}, our setup demonstrates a resolution of $\pm 0.1~\mathrm{mm}$ when converting stepper motor currents into nozzle positions.
The detail measured values, along with the corresponding true values, are presented in Table~\ref{tab:feedrate_comparison} in Appendix.

\begin{table} [!h]
  \centering  
  \caption{Position estimation error across different speed.}
\begin{tabular}{|c | c| c| c| c}
\hline
Speed(mm/min) & Min Error(mm) & Max Error(mm) & Avgerage Error(mm) \\
\hline
500  & -0.10 &  0.10 & -0.003  \\
1000 & -0.10 &  0.10 & -0.009  \\
1500 & -0.10 &  0.00 & -0.003  \\
2000 & -0.10 &  0.10 &  0.000  \\
\hline
\end{tabular}

\label{tab:feedrate_error}
\end{table}
\vspace{-15pt}

\subsection{Stepper Motor Speed Estimation}
Another important aspect of stepper motor performance is the speed at which motions are executed. As discussed in Section~\ref{sec:sterpperStopStart}, our hardware receives a signal whenever a new motion starts and when it stops. By calculating the time interval between these two events and using the known motion distance, the execution speed can be estimated.

\begin{figure}[!h]
    \centering

    \includegraphics[width=0.5\linewidth]{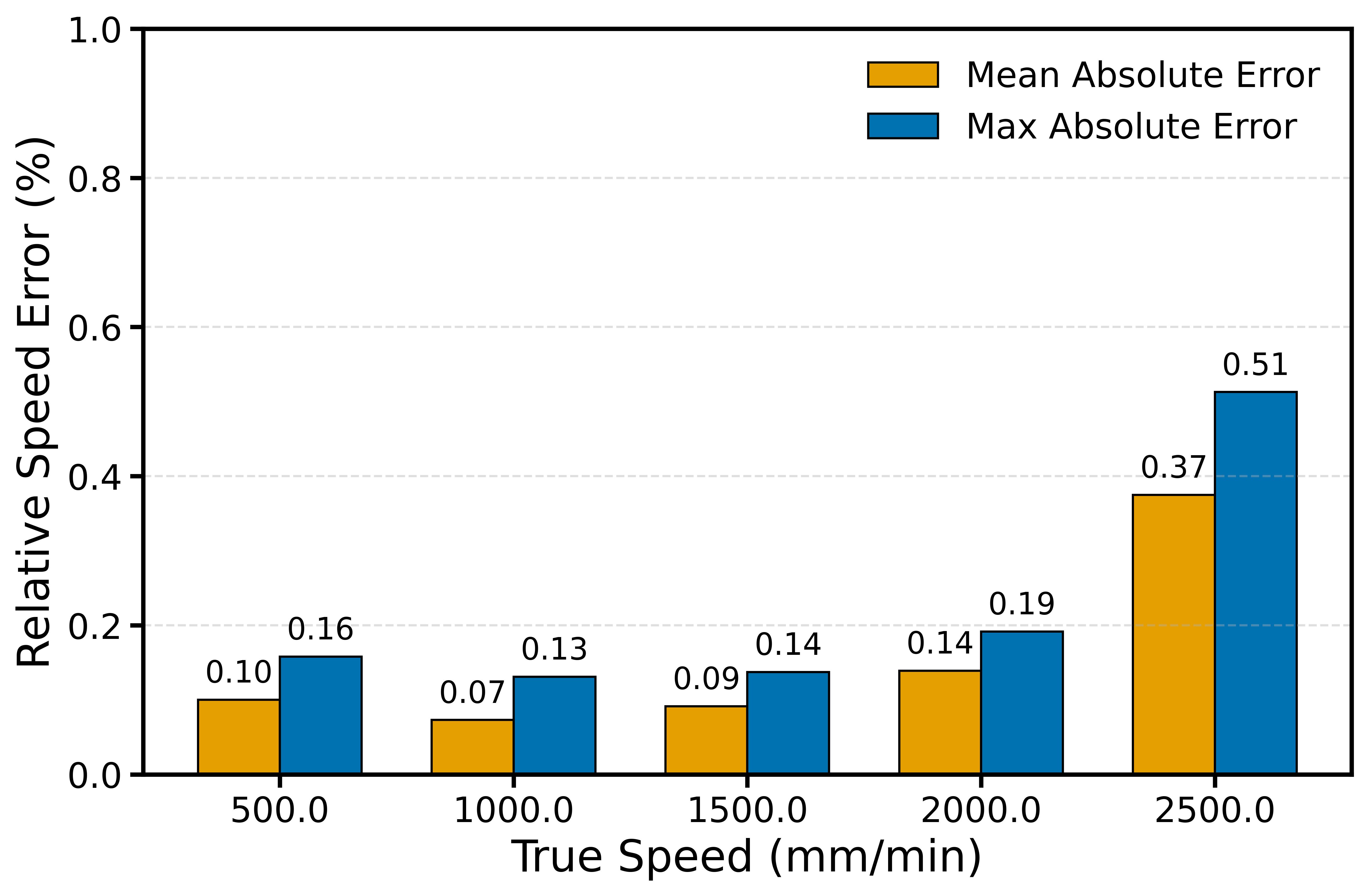}

\caption{Speed estimation errors for a motion path of length 200\,mm. The mean and maximum errors are shown for different speeds. The maximum error is less than 1\%. }
    \label{fig:stepperSpeed}
\end{figure}

To evaluate this behavior, we performed tests with constant-length motions at different speeds. The measured results are shown in Figure~\ref{fig:stepperSpeed}. For each speed cluster, we analyzed the distribution of the measured speeds and computed summary statistics, including the mean and maximum speeds. Each test was performed more than 15 times to ensure the reliability of the measurements. Our setup shows a maximum speed error of less than 1\% when the commanded speed is 2500~mm/min for a 20~mm long motion (See Figure~\ref{fig:stepperSpeed}). Further sensitivity analysis of speed measurement is also provided in the Appendix section \ref{sec:speedAnalysis}. 

\vspace{-35pt}

\section{Firmware Attacks and Defense Scenarios}
\label{sec:attackAndDefense}
We examine several attack scenarios established in prior research and demonstrate the effectiveness of our defense method in mitigating these threats. Based on our observations, firmware attacks have four main targets: motion actuators, temperature-related actuators (such as cooling fans, the heated bed, and the hotend), and stop-switch sensors.

\subsection{Motion Targeted Attacks}
3D printers have at least three stepper motors, one for each axis, and additional stepper motor is used for filament extrusion. If any of the motion stepper motors is compromised, the printer will produce faulty movement. If the extrusion stepper motor is compromised, it will affect the amount of material being extruded. We will evaluate some attacks that specifically target stepper motors. 
Print Your Own Grave Attack was proposed by \cite{rais2024sok}. In this scenario, the 3D printer's firmware is compromised in such a way that it can shatter the printer’s outer glass door. To achieve this, the printer first manufactures a rectangular object with a specifically designed hole in its center. Subsequently, the object is pushed against the outer glass with sufficient force to break it. This process consists of two phases: the first phase involves unauthorized printing of the object, while the second phase executes a command to forcefully push the printed part against the printer’s glass door. In another scenario the attacker compromises the firmware in such a way that the internal cavity of the printed object is deliberately misaligned. Instead of extending vertically as intended, the cavity is gradually tilted at a slight angle, potentially compromising the mechanical integrity and functional reliability of the part\cite{rais2021dynamic}. It is also possible to target the speed of motion. In such attacks the adversary manipulates the filament extrusion speed during printing, causing localized variations in filament width. Specifically, some regions of the printed object receive more filament, making them thicker, while other regions receive less, resulting in thinner sections\cite{rais2021dynamic,rais2022low,bayens2017see,flaw3d}.

\subsection{Defense Mechanisms Against Motion Targeted Attacks}
Based on the provided attack scenarios that compromise motion execution, two primary parameters are susceptible to compromise: G-code length and G-code velocity. For instance, in the length-based attack scenario, a legitimate motion command specifies a displacement of $L$, while the executed motion deviates by a perturbation $\Delta L$, resulting in an actual displacement of $L \pm \Delta L$. Conversely, in the speed-based attack scenario, the intended motion executes at velocity $v$, while the actual execution speed is modified to $v \pm \Delta v$, where the commanded motion is executed at speed $v$ but the actual command is executed at speed $v \pm \Delta v$. To validate these defense scenarios, we present two representative test cases in the following sections.
\vspace{-10pt}

\subsubsection{Defense for  Motion Attacks:}
To illustrate this scenario, we created a custom G-code file in which the nozzle head traverses a square path of 30~mm by 30~mm. In the first scenario, we executed the normal G-code and recorded both the decoded results and the stepper motor data using our Firewall3D hardware.  

\begin{figure*}[!h]
    \centering
    \includegraphics[width=1\linewidth]{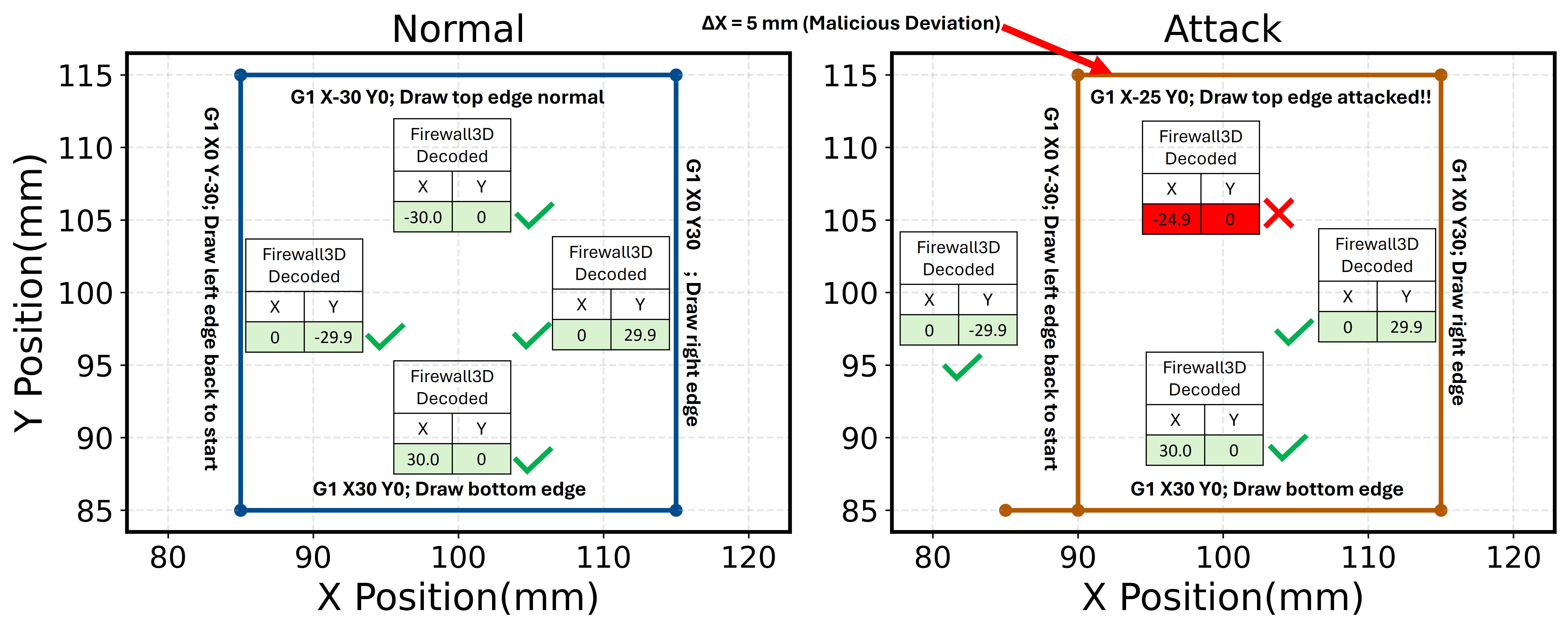}
    \vspace{-10pt}
    \caption{Comparison of normal and attacked motion execution. The left plot shows correct 30~mm movements, while the right plot highlights a compromised command with a 25~mm deviation detected by Firewall3D. Firewall3D expects a motion of 30~mm; however, only 25~mm of motion is executed.}
    \label{fig:stepperAttackLocation}
\end{figure*}

In the second scenario, we assumed that a single line of the G-code commands was compromised (see Figure~\ref{fig:stepperAttackLocation}), causing the top edge of the square to be drawn incorrectly. Instead of a 30~mm motion, a 25~mm motion was executed, resulting in a deformation of the shape.  

As shown in the normal scenario, the output of the Firewall3D circuit  matches the expected and executed commands. In contrast, in the attacked scenario, Firewall3D can easily detect the inconsistency between the intended 30~mm command and the executed 25~mm command, identifying the 5~mm discrepancy.

\vspace{-10pt}
\subsubsection{Defense for Speed Based  Attacks}
To prove that our hardware can successfully mitigate attacks that manipulate the speed of the stepper motors, we conducted an experiment. First, we drew the same square shown in Figure~\ref{fig:stepperAttackLocation}, with the speed of 1500~mm/min for all edges in the normal scenario.

\begin{figure}[!t]
    \centering
    \includegraphics[width=0.65\linewidth]{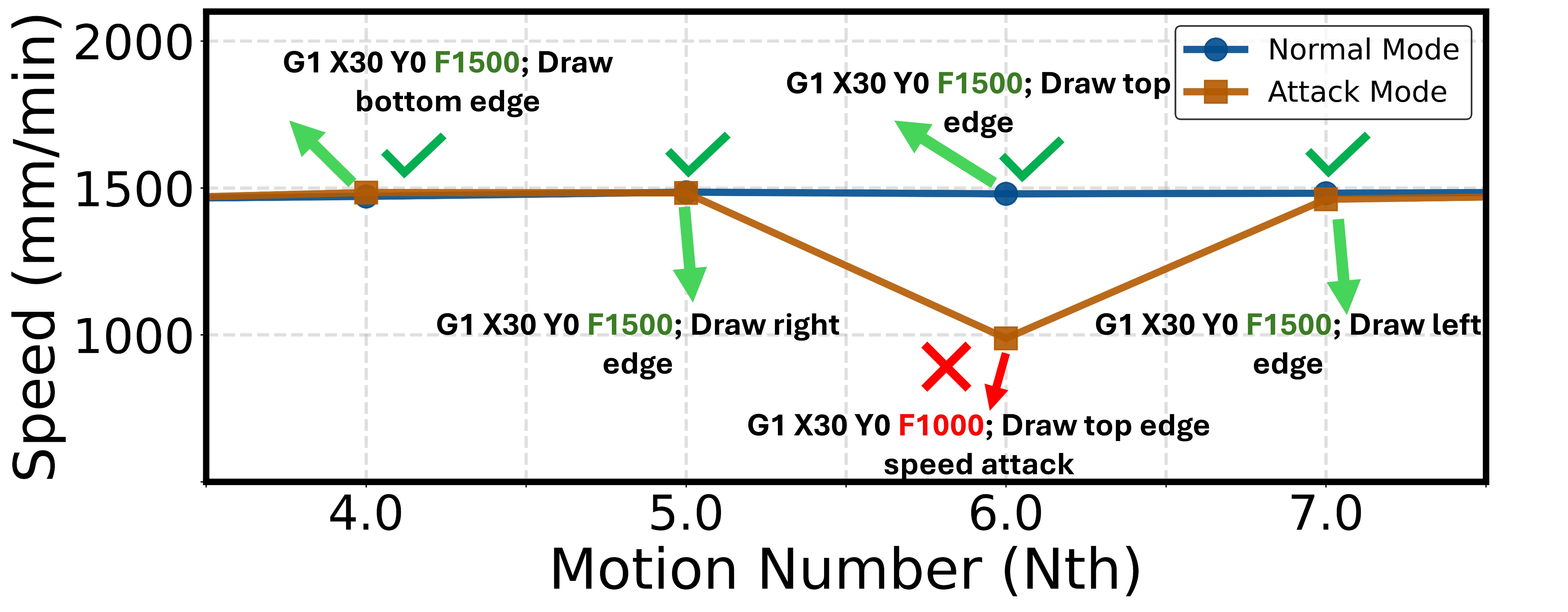}
    \caption{Comparison of 3D printer normal operation and under attack. Normal Mode maintains consistent speed (~1500 mm/min) across all four edges, while Attack Mode shows the drop at Motion speed for the top edge drawing, with speed dropping to 1000 mm/min. Firewall3D can detect the anomaly.}
    \label{fig:stepperAttackspeed}
\end{figure}

In contrast, in the attack scenario, the attacker manipulates the speed such that one of the edges is created with 1000~mm/min instead of 1500~mm/min. As shown in Figure~\ref{fig:stepperAttackspeed}, the deviation between the true and manipulated values is clear, and our setup successfully detected this speed variation.

\subsection{Temperature Targeted Attacks}

Cooling fans, the hotend nozzle, and the heated bed are key components responsible for temperature regulation in a 3D printer. We will discuss the possible firmware vulnerability that can cause serious damage.

Calibrating a 3D printer requires tuning critical parameters like filament retraction, hotend temperature, and cooling fan speed to achieve quality prints. Attackers can exploit this calibration process by dynamically altering parameters during operation, misleading operators into believing their settings are suboptimal. The compromised firmware continuously modifies values in the background, preventing discovery of truly optimal configurations~\cite{rais2024sok}.

An attacker can also set the nozzle temperature to maximum while disabling cooling fans. The filament inside the extruder melts and clogs the nozzle~\cite{gao2018watching,rais2024sok}. In addition, excessive heat can melt nearby wires, which can short circuit and potentially trigger fire hazards.

Moreover, an adversary can manipulate hotend temperature during printing to induce internal defects invisible to visual inspection. While the printed part appears geometrically sound, imposed thermal variations degrade material bonding and microstructure across layers~\cite{rais2021dynamic}.

In a timing attack scenario, compromised firmware receives a temperature command but falsely reports reaching the target (e.g., 150°C) while continuing to heat. The actual hotend may exceed 250°C undetected. Although safety mechanisms normally trigger thermal runaway protection if target temperature isn't reached within a time window, compromised firmware can disable this protection, allowing uncontrolled heating and severe damage.

Cooling fans generate distinct acoustic signatures at different speeds~\cite{asgar2026quietprint}. An attacker can modulate firmware to vary fan speed in patterns encoding sensitive design information. An insider could place an audio recorder near the printer to capture these acoustic patterns and extract confidential data.

\vspace{-10pt}
\subsection{Defense Mechanisms Against Thermal and Cooling Fan Manipulation Attacks}To detect temperature-based attacks on a 3D printer, it is first necessary to model the expected thermal behavior of the system. For the described attack scenarios  the timing characteristics of the temperature response must be formally characterized. For example, if the initial hotend temperature is 50\,°C and the G-code command sets a target temperature of 150\,°C, the system should estimate the expected time required to reach the desired setpoint under normal operating conditions. After this expected time window, the temperature should be continuously monitored to ensure that it remains within an acceptable tolerance band around the target value, with limited variance. Any significant deviation in the heating rate, overshoot behavior, stabilization time, or steady-state variance may indicate abnormal behavior. To model the timing behavior, we designed two simulation scenarios. In the first scenario, the initial nozzle temperature was set to 50\,°C, and the target temperature was increased to 200\,°C. In the second scenario, the experiment was repeated with a target temperature of 150\,°C. The acquired temperature profiles for both cases are shown in Figure~\ref{fig:temptime}. Based on the acquired values, the slope of the temperature rise is approximately constant and can be expressed as:
\begin{figure}[!b]
    \centering
    \includegraphics[width=0.6\linewidth]{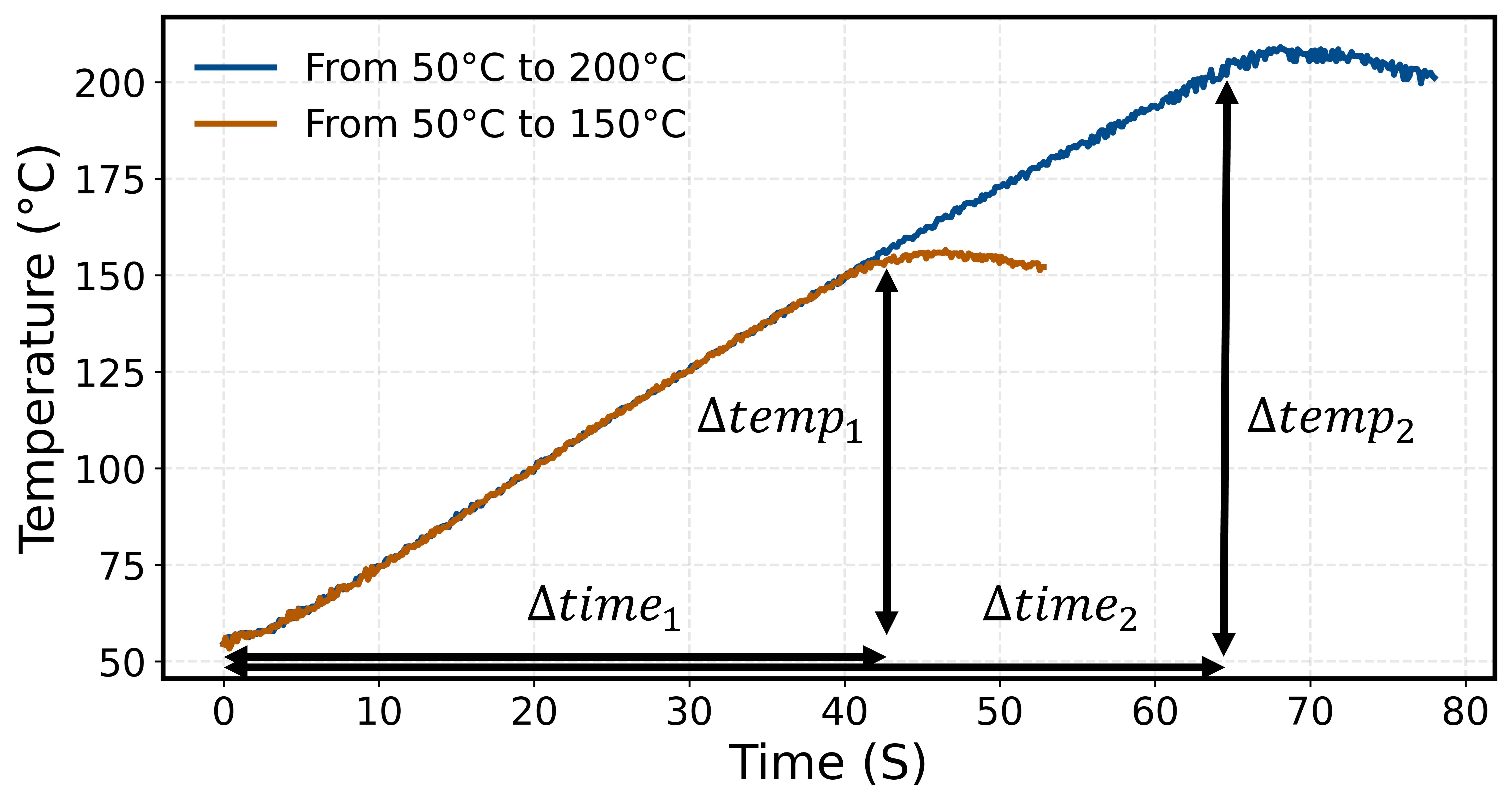}
     \caption{Transient thermal response of the nozzle showing the rise time from 50\,°C to 200\,°C and from 50\,°C to 150\,°C.}
    \label{fig:temptime}
\end{figure}
\[
\frac{\Delta Temp_1}{\Delta time_1}
= \frac{150 - 50}{43}
= \frac{100}{43}
\approx 2.32 \, ^\circ\mathrm{C}/\mathrm{s}
\]

\[
\frac{\Delta Temp_2}{\Delta time_2}
= \frac{200 - 50}{65}
= \frac{150}{65}
\approx 2.31 \, ^\circ\mathrm{C}/\mathrm{s}
\]
Although these values depend on the PID controller coefficients, once the parameters are fixed, the system is expected to demonstrate consistent and repeatable thermal behavior. As shown above, the slopes corresponding to different target temperatures exhibit similar values.

Therefore, we define a waiting time \( t_{wait} \) proportional to the temperature difference:
\vspace{-10pt}

\[
t_{\text{wait}} 
= \gamma (\, \frac{T_{\text{set}} - T_{\text{current}}}{\alpha})
\]

where \( \alpha \) denotes the estimated heating-rate coefficient, and \( \gamma \) is a safety factor used to account for settling time. after the waiting interval is passed, the measured temperature is examined to confirm that it satisfies the specified tolerance constraints. In particular, we define a deviation parameter \( \beta \), corresponding to a 10\% permissible error margin relative to the target temperature. The temperature must satisfy
\vspace{-5pt}
\[
\left| T_{\text{measured}} - T_{\text{set}} \right|
\leq \beta \, T_{\text{set}}.\]

\begin{figure*}[!t]
    \centering
    \includegraphics[width=1\linewidth]{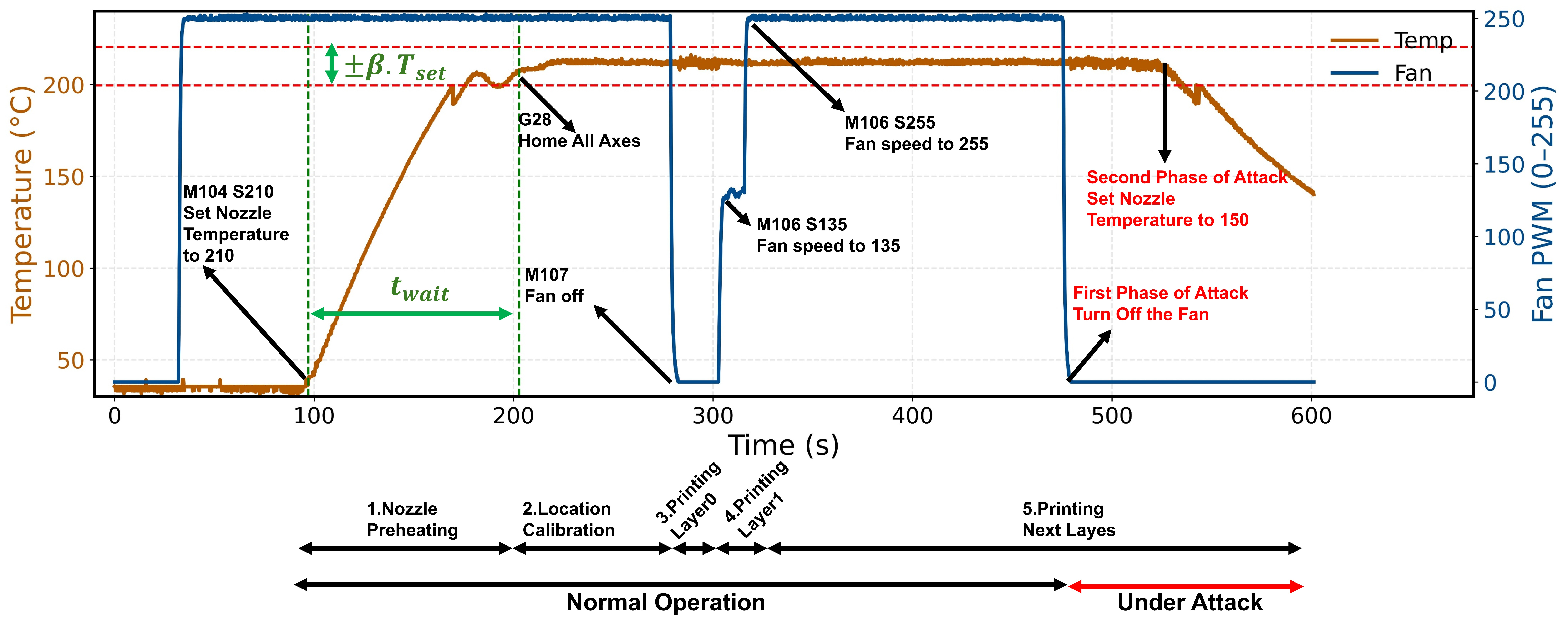}
    \vspace{-5pt}

    \caption{Temperature and cooling fan monitoring during normal operation and a firmware attack scenario. In the normal operation phase, the G-code commends as well as their effect on temperature and cooling fan is shown. During the attack phase, the temperature and fan speed goes out of the valid range. }
    \label{fig:tempattack}
\end{figure*}

Any deviation beyond these bounds is considered indicative of an anomaly. We conducted the attack scenario according to Figure~\ref{fig:tempattack}. Initially, the printer executes normal initialization commands, such as setting the nozzle temperature and calibrating the printer axes. The command \texttt{M104 S210} is issued to set the nozzle temperature to 210$^\circ$C. The corresponding waiting period and the acceptable settling-time threshold range are shown in the figure. After the nozzle temperature reaches the expected value and the calibration of all axes is completed, the printing process begins. During the first layer, the cooling fans are turned off to improve filament adhesion to the build plate. For the following layer, the printer operates at half speed, and for the remaining layers it continues at full speed, as illustrated by the blue graph in the figure. The printing process proceeds normally until the first phase of the attack is triggered.As shown in the figure, the expected nozzle temperature during printing is 210$^\circ$C. However, the infected firmware maliciously changes the temperature setpoint to 150$^\circ$C. The moment the measured temperature deviates outside the predefined threshold band, a safety alert is triggered. This allows the system to immediately stop the printing process, thereby preventing the attack from continuing. The cooling fan attack is also illustrated in the same figure. A similar threshold-based monitoring mechanism and waiting-time strategy can be applied to detect anomalies in fan behavior. Under normal conditions, the fan speed should remain at 255 (maximum speed). However, during the attack the firmware forces the fan speed to drop to 0. Once the fan speed falls outside the expected threshold range, the system detects the anomaly and raises an alert, indicating a potential attack.

\vspace{-10pt}

\subsection{EndStop Switch Attack and Defense  }
\begin{figure}[!h]
    \centering
    \includegraphics[width=0.5\linewidth]{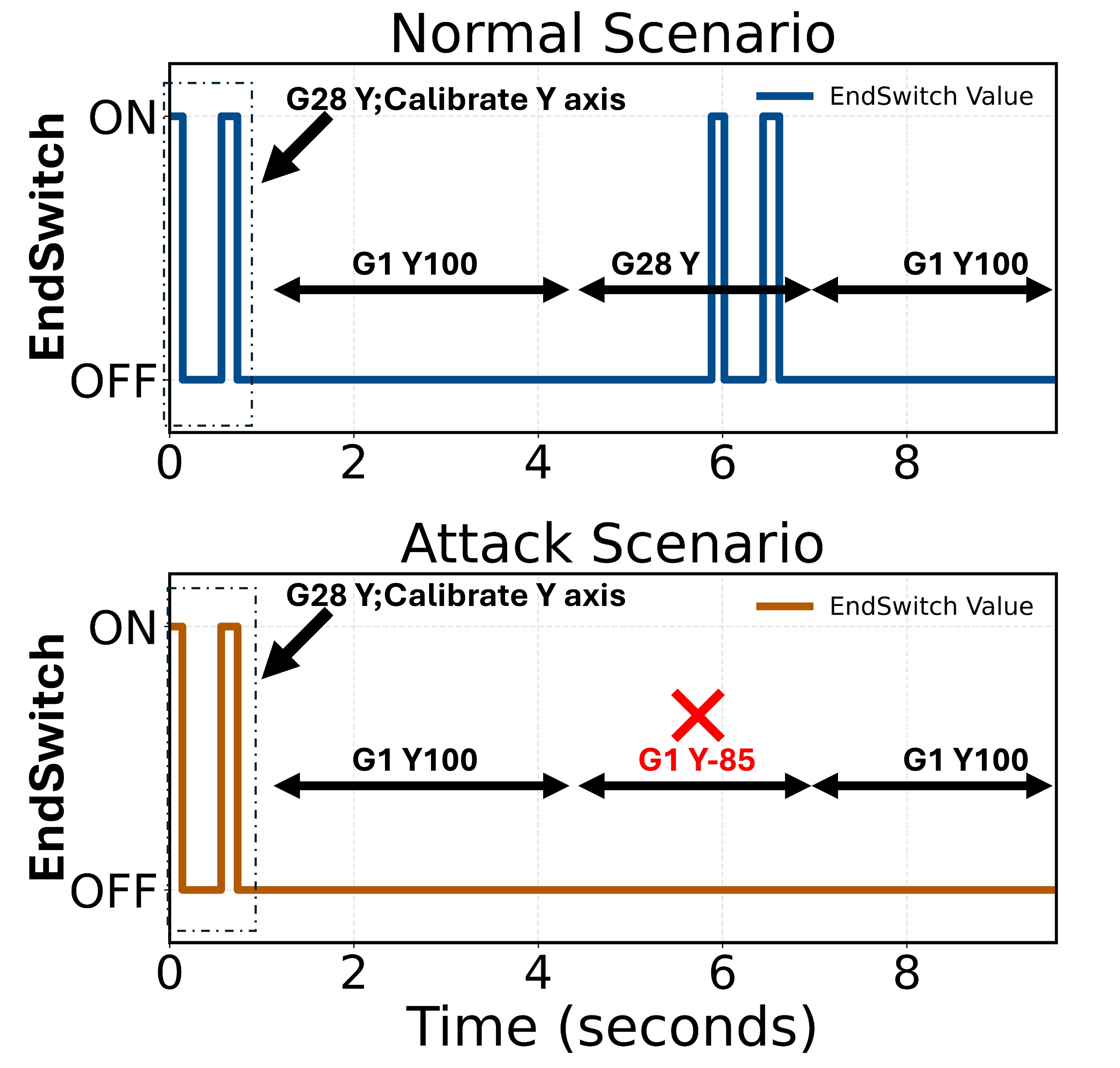}

    \caption{The Normal Scenario (top) shows standard calibration operations where the Y-axis endstop switch is triggered twice during each calibration with the (G28 Y) command and with a On-Off-On-off pattern. This axis is calibrated twice in the Normal scenario. The Attack Scenario (bottom) demonstrates the simulated attack condition (G1 Y-85) instead of the homing procedure. In the attack scenraio, the calibration pattern is only seen once as the attacker changes the calibration command with another motion(G1 y-85).Hence the Endswitch is not clicked.}
    \label{fig:stopswithcattack}
\end{figure}

As shown in Figure~\ref{fig:stopswithcattack}, we evaluate two scenarios to demonstrate the vulnerability of 3D printers to endstop manipulation attacks. In the normal scenario, the Y axis is first calibrated using the G28~Y homing command. The nozzle is then repositioned 100~mm along the Y axis. Following recalibration, the nozzle returns to the same reference position at 100~mm, ensuring accurate layer placement. In the attack scenario, after the initial calibration and 100~mm displacement, the attacker bypasses standard recalibration by moving the nozzle only 85~mm toward the endstops, deliberately avoiding trigger engagement and creating a 15~mm offset. When the subsequent motion command G1~Y100 is executed, the nozzle position deviates by 15~mm from the intended location, resulting in a layer shift throughout the remainder of the print. Since our hardware can read the endstops switches in real time, it can detect weather the switches are pressed or not. Hence, even if the attacker may compromise the firmware in such a way that it claims it already calibrated the axis, since our hardware did not detect the calibration signals, it can detect the anomaly and stop the printing process.

\section{Limitations and Future work}
In this work, we assume that firmware manipulation of the 3D printer occurs through the supply chain, faulty firmware updates, or insider threats. We do not consider an attacker capable of modifying our setup or altering the Firewall3D firmware, and we assume that the communication channel between Firewall3D and the PC performing anomaly detection is secure; however, this may not always hold in practice. As future work, a fully encrypted communication channel between Firewall3D and the PC can be implemented to prevent eavesdropping or tampering, even by advanced attackers. Additionally, firmware integrity verification mechanisms can be incorporated, allowing the PC to detect unauthorized updates or reprogramming of the Firewall3D microcontroller through this secure channel.

\section{Conclusion}
\label{sec:concl}
In this paper, we introduced Firewall3D, the first dedicated hardware firewall designed to protect 3D printers against firmware level attacks. Firmware attacks represent a critical vulnerability in additive manufacturing systems as they operate at the hardware level and are difficult to detect without invasive debugging mechanisms such as JTAG interfaces. Our work addresses this security gap by proposing a practical defense mechanism based on real-time physical layer monitoring. Firewall3D continuously monitors stepper motor currents, temperature readings, cooling fan signals, and endstop switches to verify that the printer's actual behavior matches intended G-code execution. Notably, our approach requires only minimal wiring connections and no modifications to the printer's motherboard or firmware.
Through comprehensive experimental validation, we demonstrated that Firewall3D effectively detects firmware attacks across three primary vectors: motion manipulation, temperature-based attacks, and end-stop spoofing. Our results show sub-millimeter position tracking accuracy which enables reliable anomaly detection. In all evaluated attack scenarios, Firewall3D successfully identified deviations between expected and actual behavior, triggering immediate alerts to halt the printing process and prevent component damage, intellectual property theft, or print degradation. As 3D printing systems are increasingly deployed in critical manufacturing and aerospace applications, Firewall3D provides an essential security layer to protect against hardware level threats that were previously undetectable.

\clearpage

\clearpage

\bibliographystyle{splncs04}
\bibliography{refs}

\clearpage

\section*{Appendix}

\section*{Detecting start/stop motion}
Listing~\ref{lis:gcode_toggle} demonstrates the hardware pin toggling mechanism that signals the completion of a motion and triggers the initiation of the subsequent motion command.

\begin{center}
\begin{minipage}{0.7\linewidth}
\begin{lstlisting}[caption={G-code modification to detect start/stop motion}, label={lis:gcode_toggle}]
G1 Y-200  # Motion 1
M400      # Wait Motion 1 Finishes  
M355 S0   # Toggle Pin
G1 Y100   # Motion 2
M400      # Wait Motion 2 Finishes  
M355 S255 # Toggle Pin
\end{lstlisting}
\end{minipage}
\end{center}

\section*{Stepper Motor Current to Position}
The code shown in Algorithm~\ref{alg:update_stepper} is  used to convert stepper motor current into motion values.

\begin{algorithm}[!h]
\caption{Update Stepper Motor State}
\label{alg:update_stepper}
\scriptsize
\setlength{\algorithmicindent}{1.5em}
\renewcommand{\baselinestretch}{0.9}
\begin{algorithmic}[1]

\STATE $m.angle \leftarrow \arctan2(m.B_{adc}, m.A_{adc})$
\STATE $m.angle \leftarrow m.angle \times \dfrac{180}{\pi}$

\IF{$0 \le m.angle \le 10$ \AND $m.state = DIAGONAL$}
    \STATE $m.state \leftarrow CARDINAL$
    \STATE $m.peak\_counter \leftarrow m.peak\_counter + 1$
    \STATE $m.dir \leftarrow ComputeDirection(m.prev\_angle, m.angle)$
    \STATE $m.prev\_angle \leftarrow 0$

\ELSIF{$35 \le m.angle \le 55$ \AND $m.state = CARDINAL$}
    \STATE $m.state \leftarrow DIAGONAL$
    \STATE $m.peak\_counter \leftarrow m.peak\_counter + 1$
    \STATE $m.dir \leftarrow ComputeDirection(m.prev\_angle, m.angle)$
    \STATE $m.prev\_angle \leftarrow 45 \cdot sign(m.angle)$

\ELSIF{$80 \le m.angle \le 110$ \AND $m.state = DIAGONAL$}
    \STATE $m.state \leftarrow CARDINAL$
    \STATE $m.peak\_counter \leftarrow m.peak\_counter + 1$
    \STATE $m.dir \leftarrow ComputeDirection(m.prev\_angle, m.angle)$
    \STATE $m.prev\_angle \leftarrow 90 \cdot sign(m.angle)$

\ELSIF{$125 \le m.angle \le 145$ \AND $m.state = CARDINAL$}
    \STATE $m.state \leftarrow DIAGONAL$
    \STATE $m.peak\_counter \leftarrow m.peak\_counter + 1$
    \STATE $m.dir \leftarrow ComputeDirection(m.prev\_angle, m.angle)$
    \STATE $m.prev\_angle \leftarrow 135 \cdot sign(m.angle)$

\ELSIF{$|m.angle| > 170$ \AND $m.state = DIAGONAL$}
    \STATE $m.state \leftarrow CARDINAL$
    \STATE $m.peak\_counter \leftarrow m.peak\_counter + 1$
    \STATE $m.dir \leftarrow ComputeDirection(m.prev\_angle, m.angle)$
    \STATE $m.prev\_angle \leftarrow 180$
\ENDIF

\end{algorithmic}
\end{algorithm}

\section*{Cooling Fan Circuit Simulation}
\label{sec:coolfansimu}
\begin{figure}[!h]
    \centering
    \includegraphics[width=0.8\linewidth]{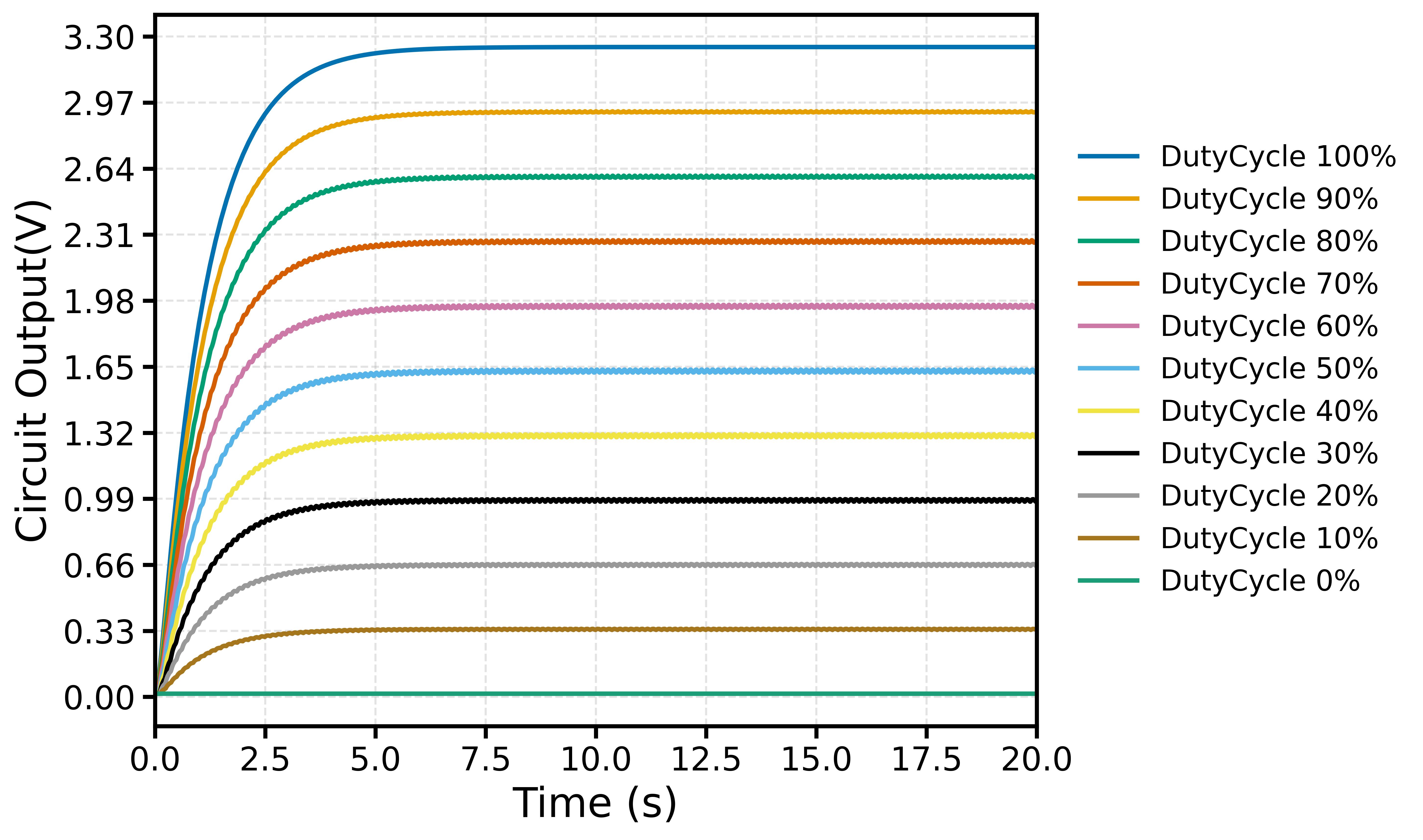}
    \caption{Simulation of the proposed circuit for converting duty-cycle values into a DC voltage for fan-speed estimation.  }
    \label{fig:fansimulation}
\end{figure}

As the simulation results are shown in Figure~\ref{fig:fansimulation}, we tested 10 linearly spaced duty cycle values (0\%,10\%,20\%, ...,100\% ). The output voltage of the circuit also varies linearly, with a voltage difference of approximately 330~mV between each 10\% increment in the duty cycle. The circuit rise time latency is also less than 5~s. Therefore, the output of our circuit exhibits a linear relationship with the duty cycle of the fan. 

\clearpage

\section*{Length Estimation Error }

Table \ref{tab:feedrate_comparison} provides the true motion and the measured value at various speeds.

\begin{table}[!h]
\centering
\scriptsize
\setlength{\tabcolsep}{4pt}
\renewcommand{\arraystretch}{1.2}
\caption{Measured and true motion values, along with the corresponding errors, at speeds of 500, 1000, 1500, and 2000 mm/min. All error values are reported in millimeters.}\label{tab:feedrate_comparison}
\begin{tabular}{|c | c c | c c | c c | c c|}
\hline
Motion(mm) &
\multicolumn{2}{c|}{500(mm/min)} &
\multicolumn{2}{c|}{1000(mm/min)} &
\multicolumn{2}{c|}{1500(mm/min)} &
\multicolumn{2}{c|}{2000(mm/min)} \\
 &
 Meas. & Err. &
 Meas. & Err. &
 Meas. & Err. &
 Meas. & Err. \\
\hline
1   & 1.0 & 0.0  & 1.0 & 0.0  & 1.0 & 0.0  & 1.0 & 0.0 \\
2   & 2.0 & 0.0  & 1.9 & -0.1 & 1.9 & -0.1 & 2.0 & 0.0 \\
1   & 1.0 & 0.0  & 1.1 & +0.1 & 1.0 & 0.0  & 1.0 & 0.0 \\
10  & 10.0 & 0.0 & 10.0 & 0.0 & 10.0 & 0.0 & 10.0 & 0.0 \\
20  & 20.0 & 0.0 & 19.9 & -0.1 & 20.0 & 0.0 & 19.9 & -0.1 \\
10  & 10.0 & 0.0 & 10.1 & +0.1 & 10.0 & 0.0 & 10.1 & +0.1 \\
20  & 20.0 & 0.0 & 19.9 & -0.1 & 20.0 & 0.0 & 20.0 & 0.0 \\
40  & 39.9 & -0.1 & 40.0 & 0.0 & 40.0 & 0.0 & 39.9 & -0.1 \\
20  & 20.1 & +0.1 & 20.1 & +0.1 & 20.0 & 0.0 & 20.1 & +0.1 \\
30  & 30.0 & 0.0 & 30.0 & 0.0 & 30.0 & 0.0 & 30.0 & 0.0 \\
60  & 59.9 & -0.1 & 59.9 & -0.1 & 60.0 & 0.0 & 59.9 & -0.1 \\
30  & 30.1 & +0.1 & 30.0 & 0.0 & 30.0 & 0.0 & 30.1 & +0.1 \\
40  & 40.0 & 0.0 & 40.1 & +0.1 & 40.0 & 0.0 & 40.0 & 0.0 \\
80  & 80.0 & 0.0 & 79.9 & -0.1 & 80.0 & 0.0 & 79.9 & -0.1 \\
40  & 40.0 & 0.0 & 40.0 & 0.0 & 40.0 & 0.0 & 40.1 & +0.1 \\
50  & 50.0 & 0.0 & 50.0 & 0.0 & 50.0 & 0.0 & 50.0 & 0.0 \\
100 & 99.9 & -0.1 & 100.0 & 0.0 & 100.0 & 0.0 & 99.9 & -0.1 \\
50  & 50.1 & +0.1 & 50.0 & 0.0 & 50.0 & 0.0 & 50.1 & +0.1 \\
60  & 60.0 & 0.0 & 60.0 & 0.0 & 60.0 & 0.0 & 60.0 & 0.0 \\
120 & 119.9 & -0.1 & 120.0 & 0.0 & 120.0 & 0.0 & 119.9 & -0.1 \\
60  & 60.1 & +0.1 & 60.0 & 0.0 & 60.0 & 0.0 & 60.1 & +0.1 \\
70  & 70.0 & 0.0 & 70.1 & +0.1 & 70.0 & 0.0 & 70.0 & 0.0 \\
140 & 139.9 & -0.1 & 139.9 & -0.1 & 140.0 & 0.0 & 140.0 & 0.0 \\
70  & 70.1 & +0.1 & 70.1 & +0.1 & 70.0 & 0.0 & 70.0 & 0.0 \\
80  & 80.0 & 0.0 & 79.9 & -0.1 & 80.0 & 0.0 & 80.0 & 0.0 \\
160 & 159.9 & -0.1 & 160.0 & 0.0 & 160.0 & 0.0 & 159.9 & -0.1 \\
80  & 80.1 & +0.1 & 80.0 & 0.0 & 80.0 & 0.0 & 80.1 & +0.1 \\
90  & 90.0 & 0.0 & 90.0 & 0.0 & 90.0 & 0.0 & 90.0 & 0.0 \\
180 & 179.9 & -0.1 & 180.0 & 0.0 & 180.0 & 0.0 & 179.9 & -0.1 \\
90  & 90.1 & +0.1 & 89.9 & -0.1 & 90.0 & 0.0 & 90.1 & +0.1 \\
100 & 100.0 & 0.0 & 100.0 & 0.0 & 100.0 & 0.0 & 100.0 & 0.0 \\
200 & 199.9 & -0.1 & 199.9 & -0.1 & 200.0 & 0.0 & 199.9 & -0.1 \\
100 & 100.0 & 0.0 & 100.0 & 0.0 & 100.0 & 0.0 & 100.1 & +0.1 \\
\hline
\end{tabular}
\end{table}

\clearpage

\section*{Motion Length Distribution}
 
\begin{figure}[!h]
    \centering
    \includegraphics[width=1\linewidth]{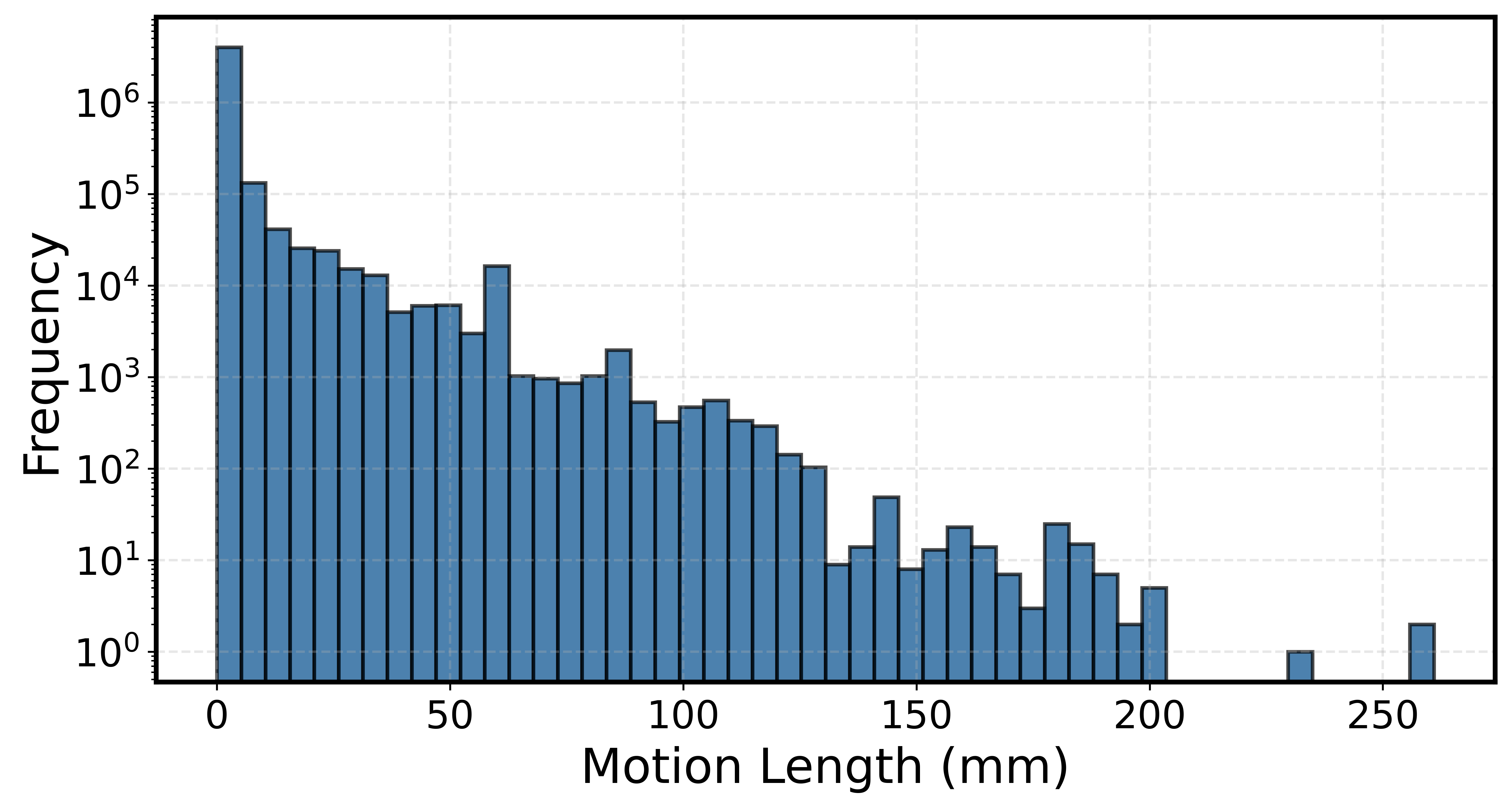}
    \caption{Distribution of G-code motion lengths on logarithmic scale. The histogram shows 4.29 million motion commands from 80 different G-code files.}
    \label{fig:gcode_motion_distribution}
\end{figure}

We analyzed the distribution of motion lengths across 4.29 million G-code commands with non-zero displacements. Motion lengths ranged from 0.001~mm to 261~mm with a mean of 2.17~mm. The distribution shows a strong skew toward small motions: 93.09\% of commands were below 5.22~mm, while only 0.91\% exceeded 10.44~mm (See Figure~\ref{fig:gcode_motion_distribution}). A secondary mode appeared around 62.64~mm, likely corresponding to specific machining operations. The histogram with logarithmic scaling reveals the long-tailed nature of the distribution, indicating that most G-code operations involve fine, short-range movements mostly because of change in the infilling process with sparse large displacements.

\section*{Speed Sensitivity Analysis}
\label{sec:speedAnalysis}
An important aspect of reporting error values and evaluating a system's measurements is to contextualize these numbers. For example, is a 3.83~mm/min speed error significant or negligible? To address this, we performed a sensitivity analysis of the speed measurement. We assumed that the timing measurements are perfect and considered only the effect of the length measurement resolution, as described in our previous algorithm. With a resolution of 0.5~mm, we estimated how deviations in the measured path length affect the calculated speed. For instance, if the actual motion length is 25~mm but the measurement yields 24.5~mm due to resolution limits, we can determine how much the resulting speed estimate would vary.

\noindent The ground-truth travel time for a given length $\Delta x$ is
\begin{equation}
\Delta t_{\mathrm{ground}} = \frac{\Delta x}{v}.
\end{equation}

\noindent Considering the spatial resolution error $\delta x$, the inferred speed is
\begin{equation}
v_{\mathrm{sensitivity}} = \frac{\Delta x - \delta x}{\Delta t_{\mathrm{ground}}}.
\end{equation}

\noindent Substituting $\Delta t_{\mathrm{ground}}$ gives
\begin{equation}
v_{\mathrm{sensitivity}} = \frac{\Delta x - \delta x}{\frac{\Delta x}{v}} = v \left(\frac{\Delta x - \delta x}{\Delta x}\right).
\end{equation}

\noindent Finally, the speed error due to spatial resolution is
\begin{equation}
e_v = v - v_{\mathrm{sensitivity}} = v - v\left(\frac{\Delta x - \delta x}{\Delta x}\right) = v \frac{\delta x}{\Delta x}.
\end{equation}

As an example if we consider a true speed of $v = 2000$ units, a measured length $\Delta x = 25$ mm, and a resolution $\delta x = 0.5$ mm. The speed error due to resolution is calculated as

\begin{align}
e_v &= v \frac{\delta x}{\Delta x} = 2000 \cdot \frac{0.5}{200} = 2000 \cdot 0.0025 = 5
\end{align}

Thus, the speed error is
\[
\boxed{e_v = 5 \text{ units}}
\]

\begin{figure}[!h]
    \centering
    \includegraphics[width=1\linewidth]{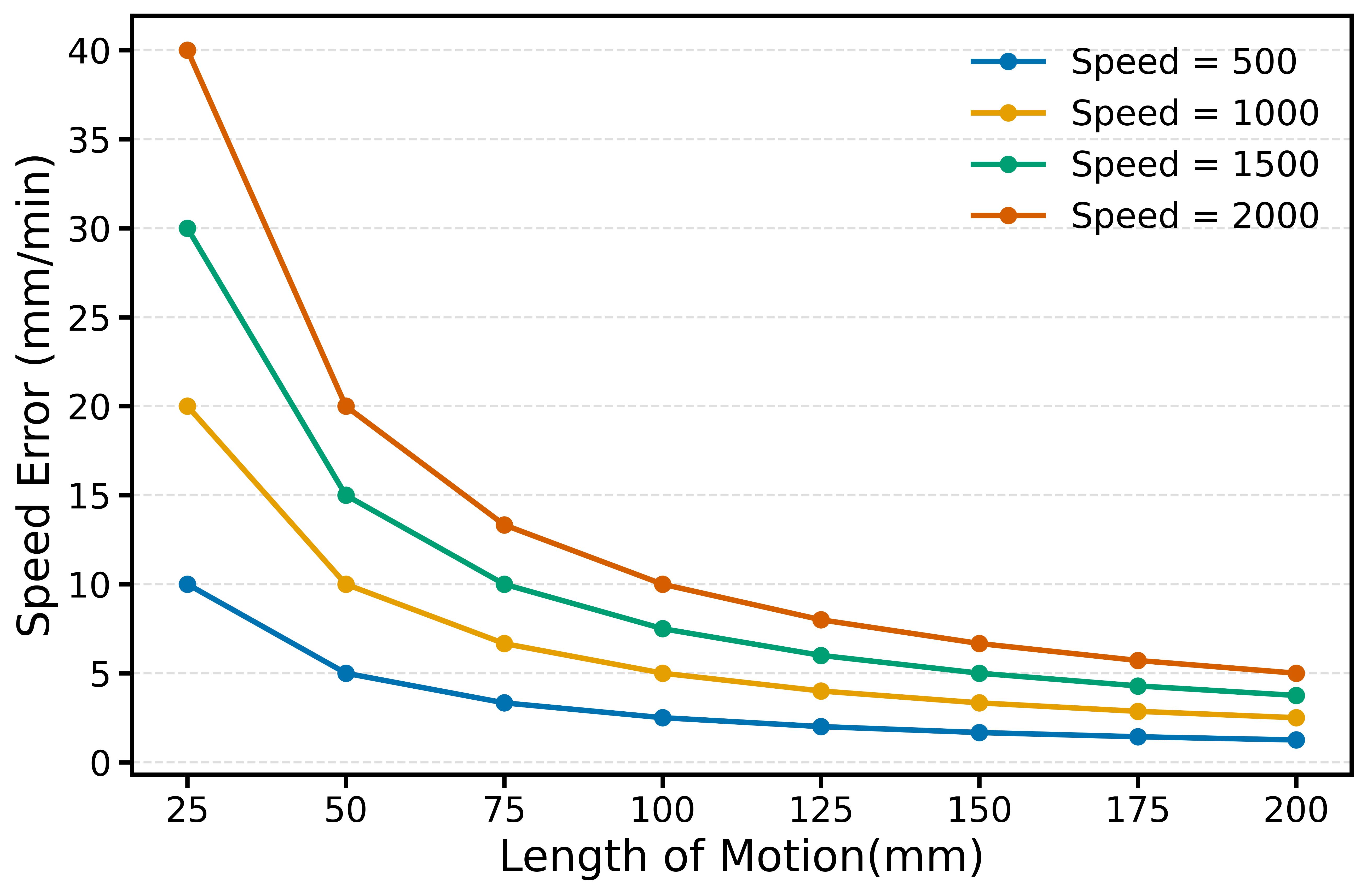}
\caption{Sensitivity of the measured speed to variations in path length. The plot illustrates how deviations in the motion length, given a 0.5~mm measurement resolution, affect the calculated speed.}
\label{fig:speed_sensitivity_vs_length}
\end{figure}

The corresponding measured speed is
\begin{align}
v_{\mathrm{sensitivity}} &= v - e_v = 2000 - 5 = 1995
\end{align}
Therefore, a maximum speed measurement error of 3.83~mm/min at a commanded speed of 2000~mm/min is within the acceptable range, as our sensitivity analysis predicts a maximum possible error of 5~mm/min, while our setup shows a lower observed error.The analytical error sensitivity plot is provided in Figure~\ref{fig:speed_sensitivity_vs_length}.

\begin{figure}[!h]
    \centering
    \includegraphics[width=1\linewidth]{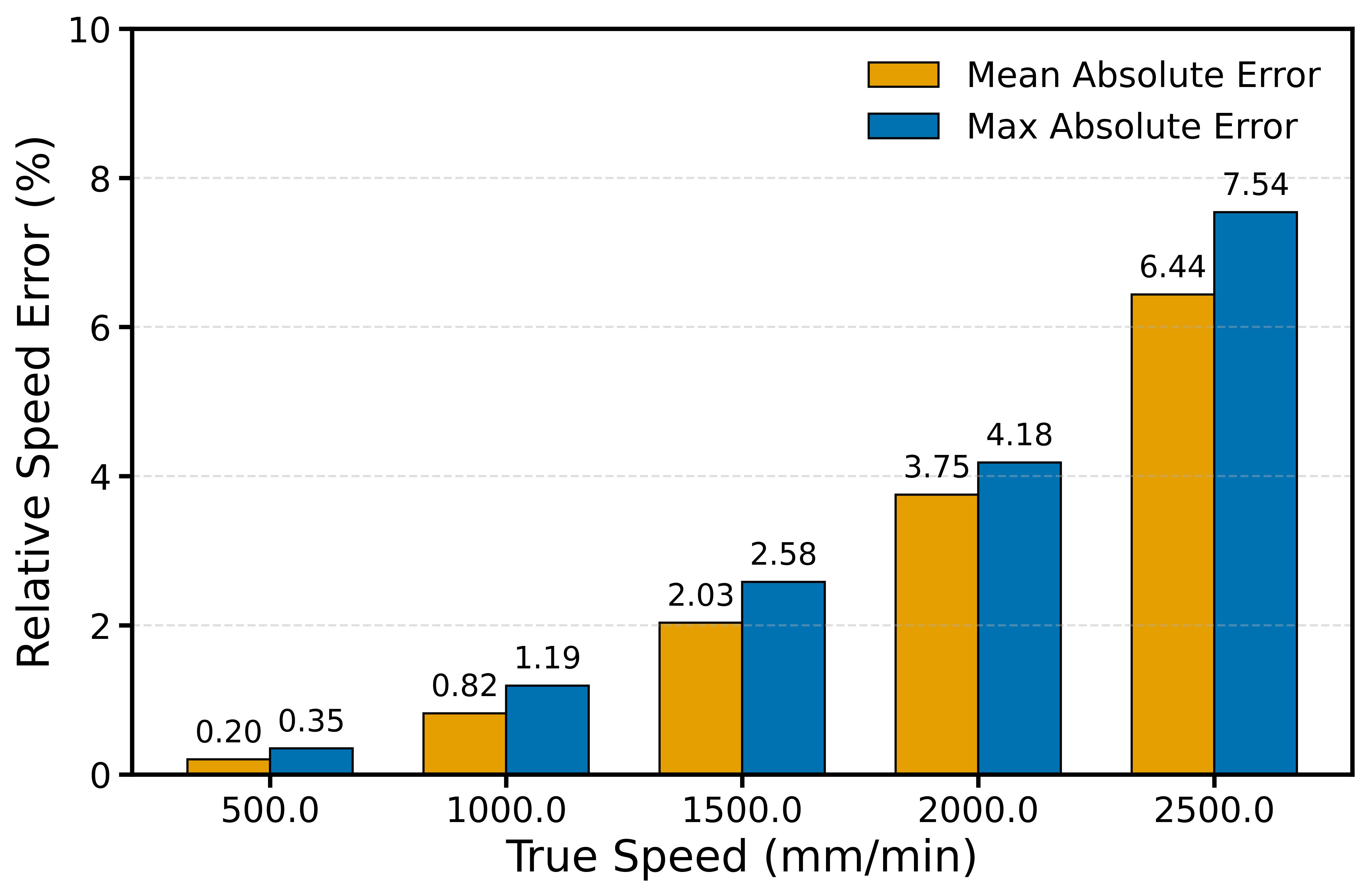}
\caption{Speed estimation errors for a motion path of length 25\,mm. The mean and maximum errors are shown for different speeds. }
    \label{fig:stepperSpeed2}
\end{figure}

\end{document}